\documentclass[sn-mathphys, iicol]{sn-jnl_mod}

\usepackage{xr}
\begin{document}
	
	\title[Classification by non-Markovian motion]{Data-driven classification of individual cells by their non-Markovian motion}
	
	
	\author[1]{\fnm{Anton} \sur{Klimek}}
	
	\author[2]{\fnm{Debasmita} \sur{Mondal}}
	
	\author[4]{\fnm{Stephan} \sur{Block}}

	\author[2,3]{\fnm{Prerna} \sur{Sharma}}
	
	\author[1,2]{\fnm{Roland R.} \sur{Netz}}

	\affil[1]{\orgdiv{Fachbereich Physik}, \orgname{Freie Universit\"at Berlin}, \orgaddress{\postcode{14195}, \state{Berlin}, \country{Germany}}}
	
	\affil[2]{\orgdiv{Department of Physics}, \orgname{Indian Institute of Science}, \city{Bangalore}, \orgaddress{\state{Karnataka}, \postcode{560012}, \country{India}}}
	
	\affil[3]{\orgdiv{Department of  Department of Bioengineering}, \orgname{Indian Institute of Science}, \city{Bangalore}, \orgaddress{\state{Karnataka}, \postcode{560012}, \country{India}}}

	\affil[4]{\orgdiv{Institut für Chemie und Biochemie}, \orgname{Freie Universit\"at Berlin}, \orgaddress{\postcode{14195}, \state{Berlin}, \country{Germany}}}

	
	
	\abstract{We present a method to differentiate  organisms solely by their motion based on the generalized Langevin equation (GLE) and use it to distinguish two different swimming modes of strongly confined  unicellular microalgae \textit{Chlamydomonas reinhardtii} (CR). The GLE is the most general model for active or passive motion of organisms and particles and in particular includes non-Markovian effects, i.e., the trajectory memory of its past. We extract all  GLE parameters from individual cell trajectories and  perform an unbiased cluster analysis to group them  into different  classes. For the specific cell population employed in the experiments, the GLE-based assignment into the two different swimming modes works perfectly, as checked by control experiments. The classification and sorting of single cells and organisms is important in different areas, our method that is based on motion trajectories offers wide-ranging applications in biology and medicine.}




	\maketitle
	\section{Introduction}
	
Classifying  individual cells or organisms is a challenging task that has been approached in many different ways and 
has ample applications. 
Distinguishing different types of cancer cells \cite{gessert2019coloncancer_images_directly,teng2015Tcell_infilt_cancer}, foodborne pathogens \cite{kang2020foodborne_hyperspectral}, sperm cells \cite{davis1995sperm_motion}, or types of neurons \cite{zeng2017neurons_classification} are just a few examples. Different techniques have been introduced to 
distinguish and classify organisms on the multi-cell  down to the single-cell level. One approach involves markers that bind  cell-specifically \cite{zhang2020artificial_marker_membrane,sanz2019marker_specific_Bcell,teng2015Tcell_infilt_cancer}.
Since  individual cells contain unique genetic and epigenetic information, 
it is also possible  to distinguish cells by their specific DNA or RNA content.
Indeed, biotechnological advances enable  single-cell RNA sequencing \cite{saliba2014rna_seq_advances,kolodziejczyk2015rna_seq_technology}, 
which can be used in combination with machine-learning approaches 
\cite{papalexi2018rna_seq_immune,qi2020rna_seq_summary}  to efficiently distinguish  single cells. 
However, RNA sequencing as well as usage of  markers require cell perturbation  or even destruction for  data acquisition.
In many cases it is desirable to classify cells without perturbing them, 
which requires label-free techniques  such as spectroscopic approaches \cite{chen2016label_free_ai_phase,kang2020foodborne_hyperspectral} or microscopy \cite{shen2015image_ai_shape_morphology}. In this way,  cell information can be extracted almost instantaneously \cite{liu2022fluorescent_live_review}   
from  living organisms \cite{gessert2019coloncancer_images_directly}. Spectroscopic and microscopic images can be processed using machine learning to classify cells, however, 
the outcomes  can be hard to interpret and require massive  training data. 
One way to simplify the processing of cell-image data is to reduce the parameter space. This can be achieved by 
feature selection  \cite{sbalzarini2002machine} or by projection onto important parameters, for instance by  
principal component analysis (PCA) \cite{qi2020rna_seq_summary}. 
A prime feature of  mobile  cells is their positional trajectory, which is relatively easy to obtain in experiments and 
 contains hidden information on the motion-generating processes within the cell \cite{boedeker2010quantitative,pohl2017inferring,maiuri2015actin_cell,heuze2013migration_voit,liu2015confinement_voit}.
Machine-learning algorithms have been proposed to classify trajectories \cite{da2019trajectory_classification} and some have been applied to single cell trajectories \cite{sbalzarini2002machine}. These approaches mostly focus on  classification and not on the interpretation of the motion patterns. 
In order to yield a mechanistic interpretation,  some specific model is usually assumed
\cite{amselem2012stochastic,selmeczi2008cell_fly,selmeczi2005cell_fly,codling2008random_benhamou,pedersen2016connect_fly,dieterich2008anomalous_klages}, which makes the interpretation model-dependent.
The most general model that describes the motion of a particle in a complex environment and captures the stochasticity of its motion is the GLE, which has been shown to accurately  describe the motion of  different cell types \cite{gail1970diffusion_mouse_fibroblasts,wright2008prw_cancer,mitterwallner_non-markovian_2020, li2011dicty,li2008_prw_dicty}.
In fact, the GLE is not an ad-hoc model but can be derived from the underlying many-body Hamiltonian \cite{mori_transport_1965,zwanzig1961memory,cihan2022_hybrid_gle}. 
Living cells are intrinsically out of equilibrium \cite{mizuno2007nonequilibrium}, 
a fact that can be properly accounted for by the GLE used  to describe the cell motion
\cite{roland_neq_2023}.

Here, we  present a method to classify individual organisms based on the  GLE parameters extracted 
from their motion  trajectories.
We apply our methods  to experimental trajectories of individual   unicellular biflagellate algae CR  \cite{jeanneret_brief_2016}
and  find two distinct groups of swimmers,
which are illustrated  in Figs. \ref{fig_1}a and b.	
In contrast to other existing methods for cell sorting and classification, 
our method requires only trajectories as input, does not require any training of a network and avoids human bias in the selection of relevant features.
Additionally, our approach allows us to  interpret motion characteristics
in terms of simple mechanistic models derived from the GLE parameters.
In the case of CR cells, the data suggest some type of elastic coupling that presumably involves
the anchoring of the flagella \cite{jeanneret_brief_2016,dutcher2016basal,wan2016chlamy_basal_coupling}, 
as schematically depicted in Fig. \ref{fig_1}c.	
Our approach is applicable to any kind of cell or organism motility data.

	\section{Results}
	
\begin{figure*}

		\centering

		\includegraphics[width=\textwidth]{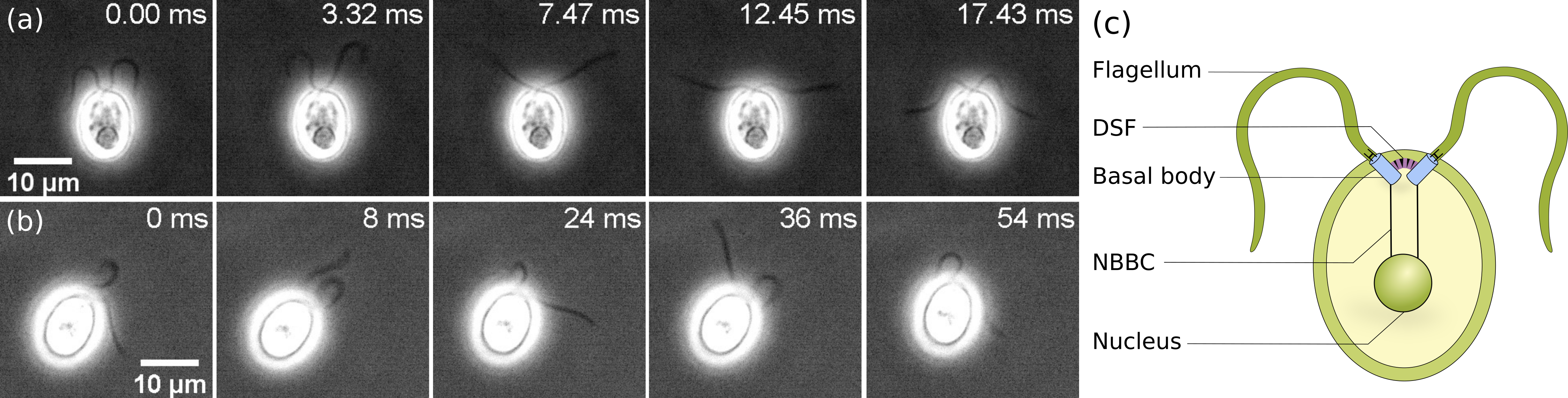}

		\caption{ Sequences of phase-contrast microscopy images of  \textit{Chlamydomonas reinhardtii} (CR) algae exhibiting
(a) synchro and (b) wobbler-type flagellar motion.  The white halo around the cells is typical for phase-contrast microscopy \cite{nguyen2017halo}. (c) Sketch of a CR cell:  the distal striated fiber (DSF) connects the two basal bodies \cite{dutcher2016basal}, which anchor the flagella and are connected to the nucleus by nuclear basal-body connectors (NBBC) \cite{jeanneret_brief_2016,wan2016chlamy_basal_coupling}.}
		\label{fig_1}
	\end{figure*}

	{\it Experimental trajectories:}
Our analysis is based on videos of $59$ CR cells that are strongly confined 
 between two glass plates separated by a distance similar to the cell diameter $\sim 10\,\mu m$, 
 which resembles the natural habitat
 of CR in soil \cite{jeanneret_brief_2016} and simplifies the recording of long
two-dimensional trajectories,  as the cells  cannot move out of the image plane of the microscope objective 
(see Methods \ref{sec_cell_preparation} and \ref{sec_recording} for experimental details).  
Videos that resolve the flagella motion  are shown in  SI Sec. I and
 reveal two different  types of flagellar motion  \cite{mondal_strong_2021}. 
In one type, the flagella move synchronously as in a breaststroke called 'synchro' (Fig. \ref{fig_1}a and Video 1), 
in the other type, the flagella move asynchronously, which results in a wobbling cell motion 
called 'wobbler' (Fig. \ref{fig_1}b and Video 2). The emergence of  two different swimming modes  reflects
 cell-size variation and constitutes the
tactile cell response  to the confining  surfaces \cite{mondal_strong_2021},
where the synchros tend to exhibit slightly larger cell bodies  and are therefore 
more strongly confined. 
Switching events between synchronous and asynchronous flagella motion are never observed and thus negligible.
We use the classification into  synchros and wobblers based on  the flagella motion 
in the high-resolution video data as a test of our 
classification  method that is based on the cell-center trajectories.

{\it Theoretical trajectory model:} 
The experiments yield  two-dimensional trajectories $x(t), y(t)$  for the cell center position, which we describe by the 
GLE equation of motion
	\begin{equation}
		\label{eq_gle}
		\ddot{x}(t) = -\int_{t_0}^{t} \Gamma_v(t-t') \dot{x}(t') dt' + F_R(t) 
	\end{equation}
with an identical equation for $y(t)$.  
Here  $\ddot{x}(t) = \dot v(t)$ denotes   the acceleration of the cell position, $\Gamma_v(t)$ is a memory kernel that 
describes how the acceleration at time $t$ depends on the cell velocity  $\dot{x}(t')= v(t')$  at previous times 
and therefore accounts for non-Markovian friction effects, and 
$F_R(t)$ is a random force that describes interactions 
with the surrounding and the interior of  the cell. 
Since the experimental system is isotropic and  homogeneous in space, no 
deterministic force term appears in the GLE. In fact,
the GLE in eq. \eqref{eq_gle}  is the most general equation of motion for an unconfined object
and  can be  derived by projection at time $t_0$  from the underlying many-body Hamiltonian  even in the presence
of non-equilibrium effects, which obviously are present for living organisms 
\cite{mori_transport_1965,zwanzig1961memory,cihan2022_hybrid_gle,roland_neq_2023}.

If the cell motion can be described as a Gaussian process,
which for CR cells  is suggested by the fact that the single-cell velocity distributions 
are perfectly Gaussian, as will be explained further below, the random force is a Gaussian process with correlations given by  
	\begin{equation}
		\label{eq_fdt_neq}
		\langle F_R(t) F_R(0) \rangle = B \Gamma_R (t) \,,
	\end{equation}
where  $B=\langle v^2 \rangle$ denotes the mean-squared cell velocity and the 
symmetric random-force kernel
is denoted as  $\Gamma_R(t)$.
In this case the equation of motion is linear and  there is  no coupling between the motion
in $x$ and $y$ direction, we thus average all cell-trajectory data over the two directions. 

For an equilibrium system, the fluctuation dissipation theorem (FDT)
predicts  $\Gamma_R(t)=\Gamma_v(\vert t\vert )$ with the mean-squared velocity given by
$B=k_BT/m$ according to the equipartition theorem,
where $m$ is the mass of the moving object and $k_BT$ denotes the thermal energy
 \cite{mori_transport_1965,zwanzig1961memory}. 
For living cells, both FDT and equipartition theorem do not hold in general
and thus there is no a priori reason why  $\Gamma_v(\vert t\vert )$ and  $\Gamma_R(t)$ should be equal
 \cite{mizuno2007nonequilibrium,netz2018non_equilibrium}. 
Nevertheless, one can construct   a surrogate model 
with an effective kernel $\Gamma(\vert t\vert) = \Gamma_R(t) = \Gamma_v(\vert t\vert)$ 
that exactly reproduces the dynamics described
by the non-equilibrium GLE with $ \Gamma_R(t) \neq \Gamma_v(\vert t\vert)$.
This can be most easily seen by considering the velocity autocorrelation function (VACF)  defined by
	\begin{equation}
		C_{vv}(t) = \langle v(0)v(t) \rangle,
		\label{eq_def_cvv}
	\end{equation}
whose Fourier transform
$\tilde C_{vv}(\omega)=\int_{-\infty}^{\infty}e^{-i\omega t} C_{vv}(t) dt$
 follows  from eqs. \eqref{eq_gle} and \eqref{eq_fdt_neq}  as
\cite{mitterwallner_non-markovian_2020}
\begin{align}
\label{eq_condition_eq_neq}
\tilde C_{vv}(\omega) = 
\frac {B \tilde{\Gamma}_R(\omega)}{( \tilde{\Gamma}_v^+(\omega) +i\omega) (\tilde{\Gamma}_v^+(-\omega) -i\omega)},
\end{align}
where $\tilde{\Gamma}_v^+(\omega)$ denotes the single-sided Fourier transform of ${\Gamma}_v(t)$
(see Methods \ref{sec_mapping_derivation} for the derivation). 
The VACF of the surrogate model with $\Gamma(\vert t\vert) = \Gamma_R(t) = \Gamma_v(\vert t\vert)$  
follows from eq. \eqref{eq_condition_eq_neq} as
\begin{align}
\label{eq_surr}
\tilde C_{vv}^{\rm sur} (\omega) 
= \frac{B \tilde{\Gamma}(\omega)}{\mid \tilde{\Gamma}_+(\omega) +i\omega \mid^2} .
\end{align}
 For each combination of 
 $\Gamma_R(t)$ and $\Gamma_v(t)$ there is a unique  $\Gamma(t)$ 
 that produces the same VACF, i.e., for which $\tilde C_{vv}^{\rm sur} (\omega) = \tilde C_{vv} (\omega)$
 holds (see SI Sec. II for the derivation) and which can 
 be uniquely extracted from trajectories via the VACF (as shown in  Methods \ref{sec_kernel_extraction}).
 Since the VACF completely determines the dynamics of a Gaussian system, as shown in Methods \ref{sec_greens_func} and in more detail in SI Sec. III,
this implies that the extracted 
 effective kernel $\Gamma(t)$  not only describes the VACF exactly but also characterizes the system completely
 \cite{roland_neq_2023b}.
 In fact, recent work, where the non-equilibrium GLE is derived from a   time-dependent Hamiltonian,
 shows that for Gaussian non-equilibrium observables the condition  $\Gamma_R(t) = \Gamma_v(\vert t\vert)$  
 is  actually  satisfied \cite{roland_neq_2023}, in line with our method  that is based on  extracting an effective kernel
 $\Gamma(t)$.

\begin{figure*}
	\centering
	\includegraphics[width=\textwidth]{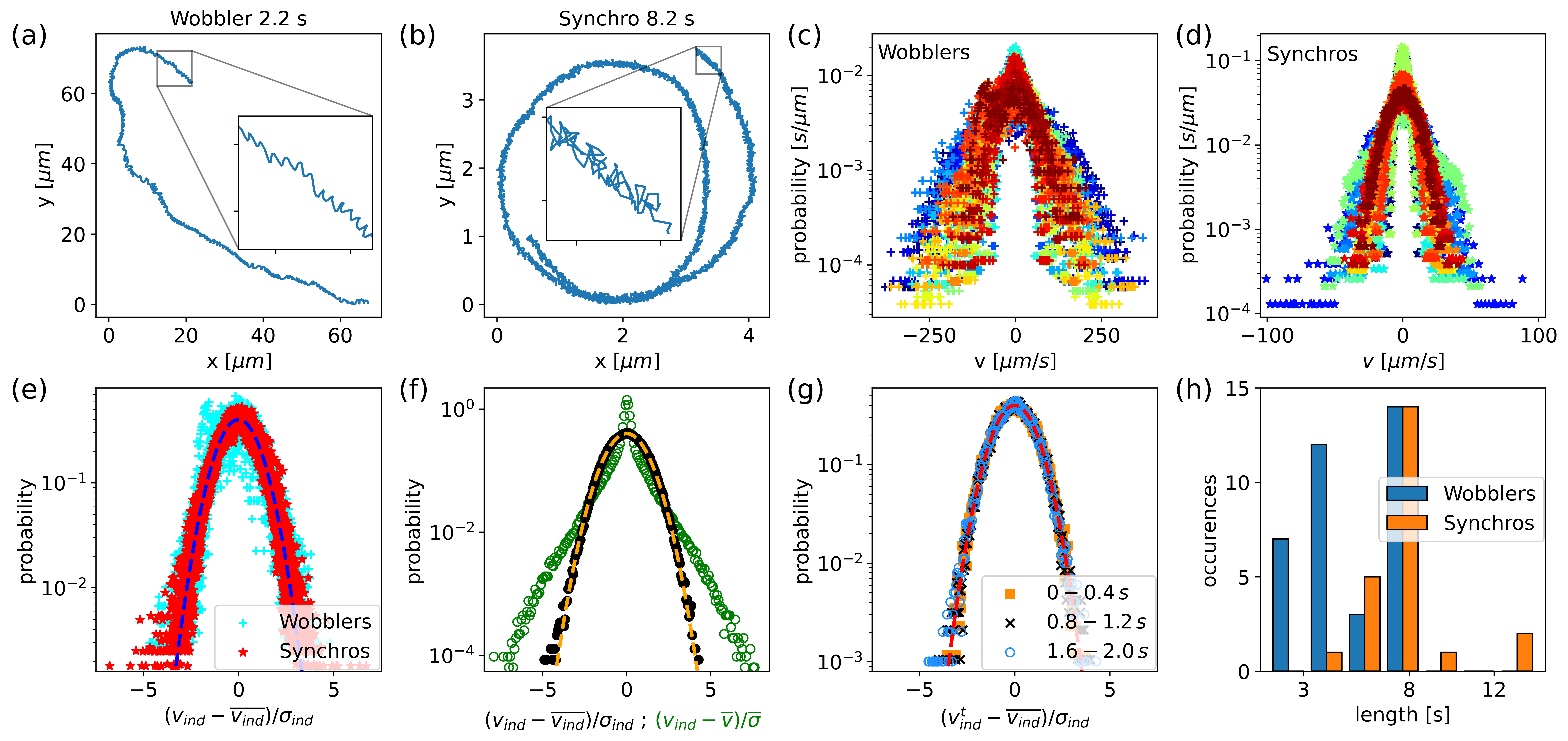}
	\caption{Exemplary cell-center trajectory \cite{Matlabtrack} (a) of a wobbler of duration $2.2\,s$ and 
		(b) of a synchro of duration $8.2\,s$. The insets show trajectory fragments of duration $0.2\,s$ each. 
		Velocity distributions of  (c) wobblers and (d) synchros,  individual cells  are distinguished by color. 
		(e) Velocity distributions of individual cells rescaled by subtracting the mean velocity of individual cells $\overline{v_{\rm{ind}}}$ and dividing by their standard deviation $\sigma_{\rm{ind}}$ for  wobblers (cyan crosses) and synchros (red stars). 
		The dashed line is the normal distribution. 
		(f) Mean velocity distribution averaged over  all cells  (wobblers and synchros). 
		For the green circles the cell velocities are rescaled  by subtracting the ensemble mean velocity $\overline{v}$ and dividing by the ensemble standard deviation $\overline{\sigma}$, for the black circles the cell velocities are rescaled  by subtracting the 
		mean velocity of individual cells $\overline{v_{\rm{ind}}}$ and dividing by their standard deviation $\sigma_{\rm{ind}}$ as in (e).
		The  normal distribution is indicated by a dashed line. 
		(g) Individually rescaled velocity distributions of all cells 
		for three different time windows. 
		(h) Distribution of recorded trajectory lengths for wobblers and synchros.}
	\label{fig_2}
\end{figure*}

{\it Trajectories and velocity distributions:}
Due to the asynchronous flagella motion of the wobblers, the cells turn 
in the flagella-beating rhythm and exhibit monotonically forward-moving  wiggly  trajectories, as shown in Fig. \ref{fig_2}a. 
In contrast, the synchronous  flagella beating of synchros leads  to fast switching between forward and backward motion, 
as shown  in Fig. \ref{fig_2}b. 
As a consequence, synchros exhibit much slower net-forward motion than wobblers, 
as seen in Figs. \ref{fig_2}a,b, where trajectories with total duration $2.2\,s$ (wobbler) and $8.2\,s$ (synchro) are compared. 
Synchros also exhibit  a somewhat narrower instantaneous  velocity distribution, 
as seen in Figs. \ref{fig_2}c,d.
However, a differentiation of the two cell types solely based on their speed does not work, as we will show later.

Even though individual cells exhibit pronounced variations in their velocity distributions, 
as seen from the large spread in Figs. \ref{fig_2}c,d, their   velocity distributions are Gaussian, 
as demonstrated  in Fig. \ref{fig_2}e: When subtracting from the cell velocities 
the mean velocity of each individual cell and dividing by the corresponding velocity standard deviation,
$(v_{\rm{ind}}-\overline{v_{\rm{ind}}})/\sigma_{\rm{ind}}$, 
all individual  velocity distributions collapse onto the Gaussian (normal) distribution (dashed line in Fig. \ref{fig_2}e).
In contrast, when subtracting from the cell velocities the  cell-ensemble mean velocity and 
dividing by the cell-ensemble standard deviation, $(v_{\rm{ind}}-\bar{v})/\sigma$, 
the velocity  distribution of all cells  deviates  strongly  from a Gaussian  (green circles in Fig. \ref{fig_2}f).
In contrast,  the individually rescaled velocity  distribution over all cells (black dots) perfectly  agrees with the Gaussian normal distribution (dashed line in Fig. \ref{fig_2}f). Thus,  single cells exhibit 
perfectly  Gaussian velocity distributions, which suggests that the GLE with Gaussian noise is  appropriate 
to analyze  experimental single-cell  trajectories.

The GLE in eq. \eqref{eq_gle} features time-independent parameters and thus describes a stationary process. 
That the cell velocity distribution does not change over the observational time is demonstrated in Fig. \ref{fig_2}g, 
where velocity distributions in three consecutive time intervals are compared 
(again subtracting the individual cell  mean velocities and dividing by the corresponding velocity 
deviations of the entire trajectory).
This suggests that the motion of individual CR algae can indeed be modeled by the GLE in  eq. \eqref{eq_gle}.
In this context, it is to be noted that wobblers are relatively fast and tend to move out of the camera window more quickly than synchros, 
leading to slightly shorter  wobbler trajectories,  as shown in Fig. \ref{fig_2}h.

\begin{figure*}		
	\centering
	\includegraphics[width=\textwidth]{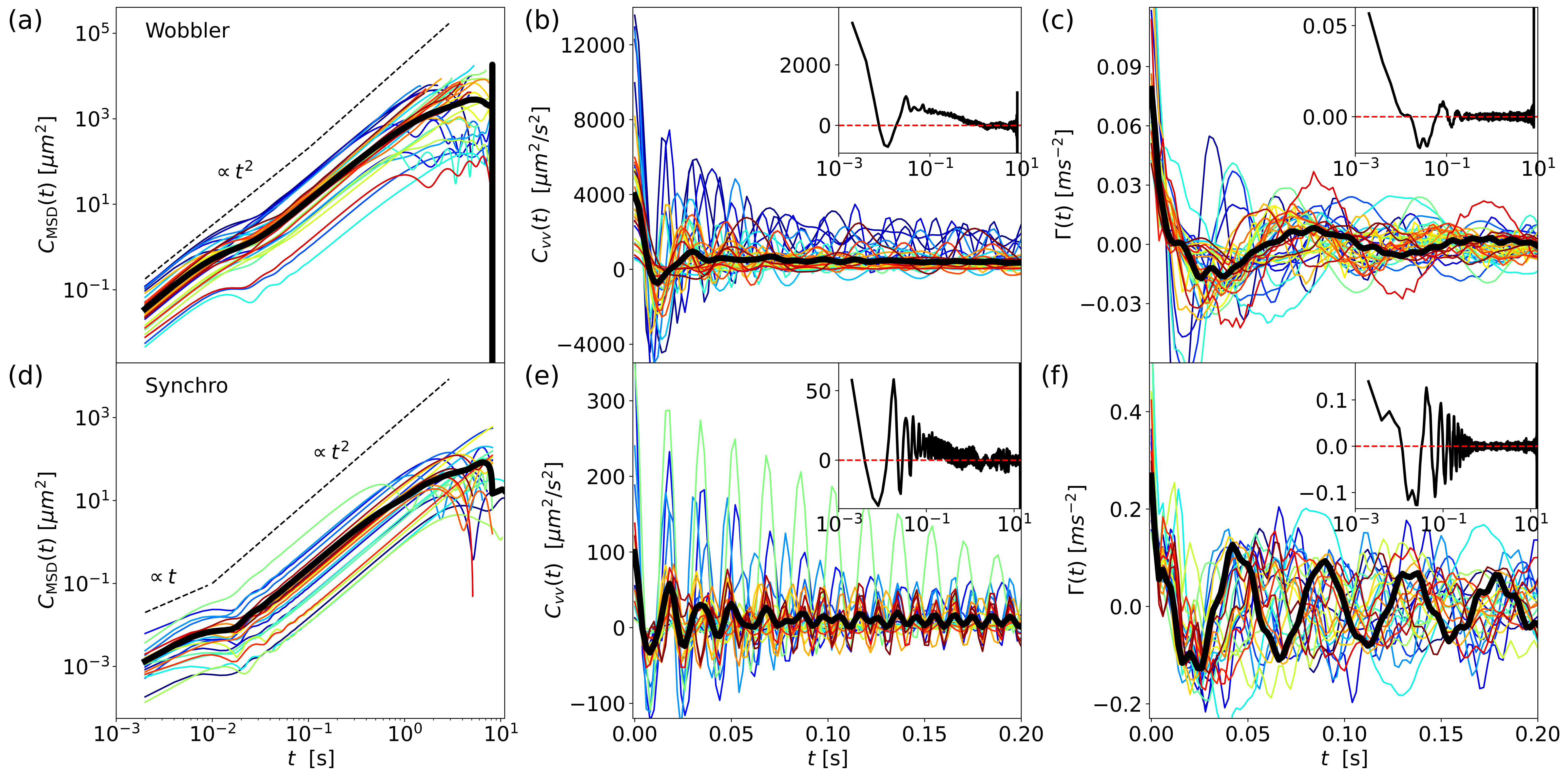}
	\caption{Results for the MSD, $C_{\textrm{MSD}}(t)$ defined in eq. \eqref{eq_def_msd}, 
		the VACF, $C_{vv}(t)$  defined in eq. \eqref{eq_def_cvv} and the friction kernel $\Gamma(t)$, extracted 
		according to  eq. \eqref{eq_G_half_step}, for wobblers in (a)-(c) and synchros in (d)-(f). 
		Different colors represent results for  individual cells, 
		the black lines in (a), (b), (d), (e) denote the average over all cells. 
		For the friction kernels the black line is computed from the average VACF. 
		The dashed lines in (a), (d)  indicate  ballistic and diffusive scaling.
		Insets  show the long-time behavior of the average quantities  on a lin-log scale.}
	\label{fig_3}
\end{figure*}

{\it  Trajectory analysis and friction kernel extraction:}
Trajectories are standardly  characterized by the VACF or   by the mean-squared displacement (MSD)
	\begin{equation}
		C_{\textrm{MSD}}(t)=\langle (x(0)-x(t))^2 \rangle\,.
		\label{eq_def_msd}
	\end{equation}
Although the  VACF is simply the curvature of the  MSD,
$C_{vv}(t) = \frac{1}{2}\frac{d^2}{dt^2}C_{\textrm{MSD}}(t)$,
the MSD and the VACF  highlight different aspects of the  trajectories.
In fact, the different propulsion modes of wobblers and synchros lead to drastically different MSDs:
The wobblers exhibit ballistic behavior $C_{\textrm{MSD}}(t) \propto t^2$
on both short and long time scales (Fig. \ref{fig_3}a) with an intermediate crossover  at $t \sim 0.04\,s$.
In contrast,  
the forward-backward motion of the synchros leads to short-time  diffusive 
behavior $C_{\textrm{MSD}}(t) \propto t$
 up to $t \sim 0.01\,s$ followed by a  long-time  ballistic regime  for $t > 0.04\,s$ (Fig. \ref{fig_3}d). 
The transition to  asymptotic diffusive behavior, expected for long times,  is  for the synchros observed 
for $t > 2s$  in extended low-resolution microscopy data,  whereas
wobblers stay in the ballistic regime for the entire observation time  of tens of seconds (see SI Sec. IV).
As  wobblers move faster than  synchros, their absolute VACF values  are higher compared to the synchros, 
as seen in Figs. \ref{fig_3}b,e. The synchros exhibit  less variation of the VACF among individual  cells, 
which leads to slowly decaying oscillations in the VACF averaged over all cells 
(black line Fig. \ref{fig_3}e compared to Fig. \ref{fig_3}b). 
These oscillations reflect   the flagella beating  cycle.

We extract  effective friction kernels $\Gamma(t)$
from the VACF of individual cells, as described in Methods \ref{sec_kernel_extraction}.
The results, shown as colored lines in Figs. \ref{fig_3}c,f, 
demonstrate  that the algal  motion deviates strongly from the simple persistent random walk model,
which is widely used to describe the motion  of organism \cite{wright2008prw_cancer,li2008_prw_dicty,gail1970diffusion_mouse_fibroblasts}
and which in the GLE formulation would correspond to $\Gamma(t)$
exhibiting a delta peak at $t=0$ and otherwise being zero.
The extracted memory kernels also reveal a substantially higher friction for  synchros compared to  wobblers, 
 in line with the fact that synchros are larger
and thus interact more strongly with the confining surfaces. 
 
Comparing Fig. \ref{fig_3}b with Fig. \ref{fig_3}c or Fig. \ref{fig_3}e with Fig. \ref{fig_3}f, one notes that  the oscillation period of the friction kernel is substantially longer than that of the VACF. 
The complex relation between the kernel and the  VACF is discussed in SI Sec. V,
where it is shown that the extracted values of the kernel decay time  and oscillation amplitude  
achieve high directionality and speed of CR cells.

\begin{figure*}
	\centering
	\includegraphics[width=\textwidth]{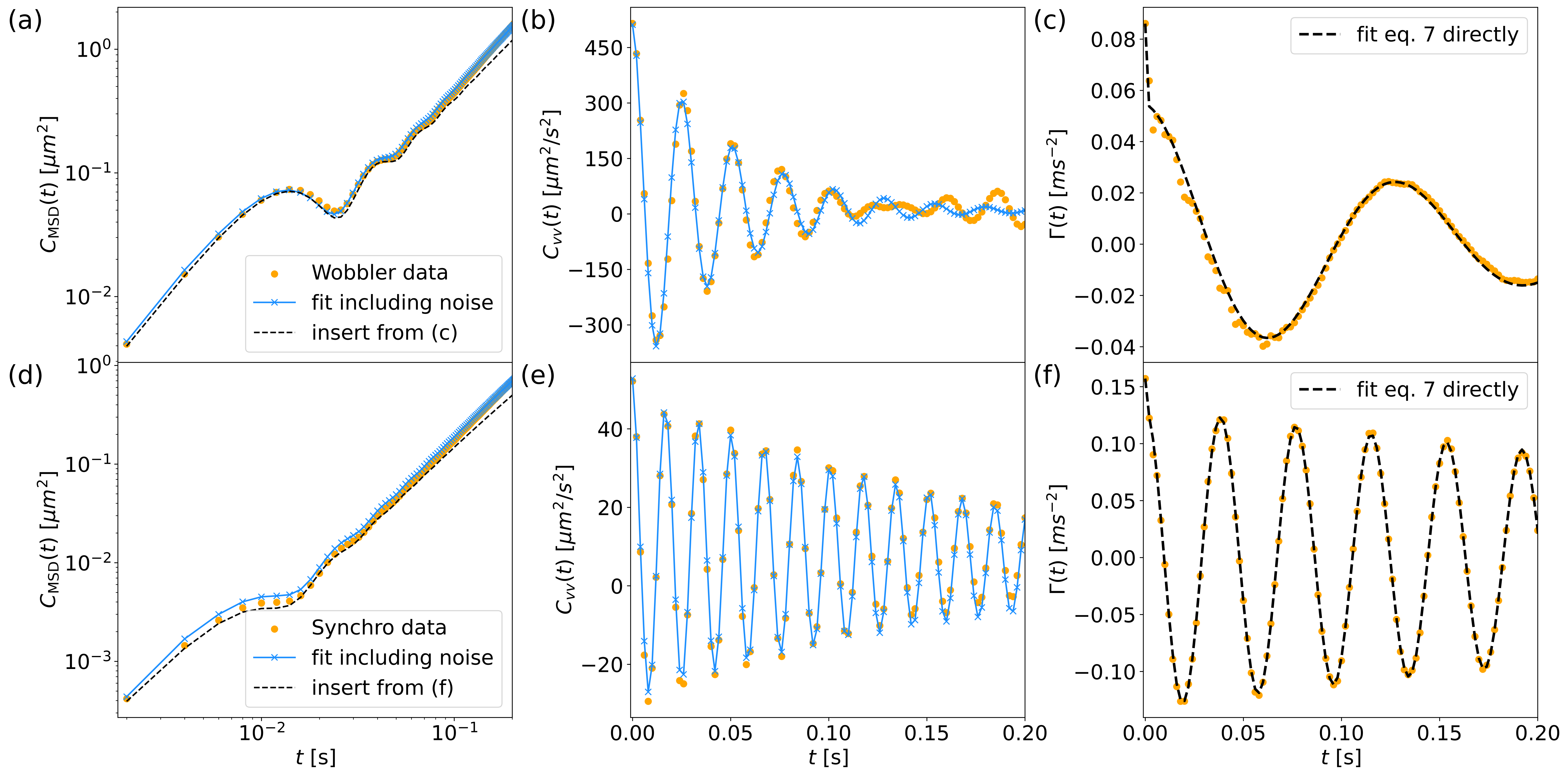}
	\caption{Results for the  MSD, $C_{\textrm{MSD}}(t)$, VACF, $C_{vv}(t)$, and friction kernel $\Gamma(t)$,  of a single (a)-(c) wobbler and (d)-(f) synchro  (orange dots). 
		The black dashed lines in (c), (f) denote fits of eq. \eqref{eq_kernel_osc_def} to the extracted friction kernel
		determining the individual cell parameters shown in Fig. \ref{fig_5}. 
		The black dashed  lines in (a), (d) denote the analytical result eq. (S51)
		obtained from the  friction-kernel fit in  (c), (f)  and the mean-squared velocity $B=C_{vv}(0)$.
		Blue crosses in (b), (e)  
		denote a fit of the discretized expression for the VACF including localization noise, eq. \eqref{eq_vacf_fit}, blue crosses in (a), (d) denote the corresponding prediction for the MSD, eq. \eqref{eq_msd_noise}, using the same parameters as in (b), (e), 
		blue crosses are connected by blue straight lines.}
	\label{fig_4}
\end{figure*}

For both wobblers and synchros,  the friction kernels exhibit an initial sharp peak followed by a decaying oscillation
and are  well described by 
	\begin{equation}
		\label{eq_kernel_osc_def}
		\Gamma(t)=2a\delta(t)+be^{-t/\tau}\cos(\Omega t)\,,
	\end{equation}
with $\delta$-peak amplitude $a$, oscillation amplitude $b$, exponential decay time $\tau$ and oscillation frequency $\Omega$. 
This is demonstrated in 
Figs. \ref{fig_4}c,f for one exemplary synchro and wobbler, where we compare the extracted memory kernels
with fits according to eq. \eqref{eq_kernel_osc_def},
see Methods \ref{sec_fitting_kernel_directly} for details. From these fits, we thus obtain four 
memory kernel parameters for each cell.

Before further analysis of the obtained individual  cell parameters, 
we test whether the GLE eq. \eqref{eq_gle} actually describes the cell motion. 
We thus  compare the experimental  MSD of a single wobbler and synchro  in Figs. \ref{fig_4}a,d (orange dots) 
with the analytical prediction based on the GLE   using 
the fit result for the friction kernel   and the mean squared velocity $B=C_{vv}(0)$
(black broken  lines, the derivation of the analytical MSD expression
is given in SI Sec. VI).  
The agreement is very good, meaning that the memory extraction works
well, which demonstrates that the  GLE is an accurate  model for the 
motion of  organisms.
 
 The MSD and VACF calculated from the GLE neglect the finite 
experimental  recording time step of $0.002\,s$, they also neglect  the localization noise of the cell position, 
due to the finite spatial resolution of the microscopy images and the projection of a three dimensional object onto a 
two-dimensional point position  \cite{mitterwallner_non-markovian_2020} (see Methods \ref{sec_fitting} for details).
The blue crosses in Figs. \ref{fig_4}b,e represent fits of the GLE-based analytical expression
for the VACF  that includes 
 localization noise and discretization effects, given in eq. \eqref{eq_vacf_fit}, to the experimental data.
  The blue crosses  in Figs. \ref{fig_4}a,d show the corresponding 
MSD results with the same  parameters according to eq. \eqref{eq_msd_noise}.
The agreement between experimental data and the discretized model is perfect, 
the fitted localization noise strength, defined in eq. \eqref{eq_msd_noise}, is of
 the order of $\sigma_{\textrm{loc}}\sim 0.02\,\mu m$, similar to the pixel size, 
 as expected (see discussion in SI Sec. VII).
 This means that  temporal and spatial discretization effects in the experimental data
 can be  straightforwardly incorporated in  the GLE model. 

\begin{figure*}
	\centering
	\includegraphics[width=\textwidth]{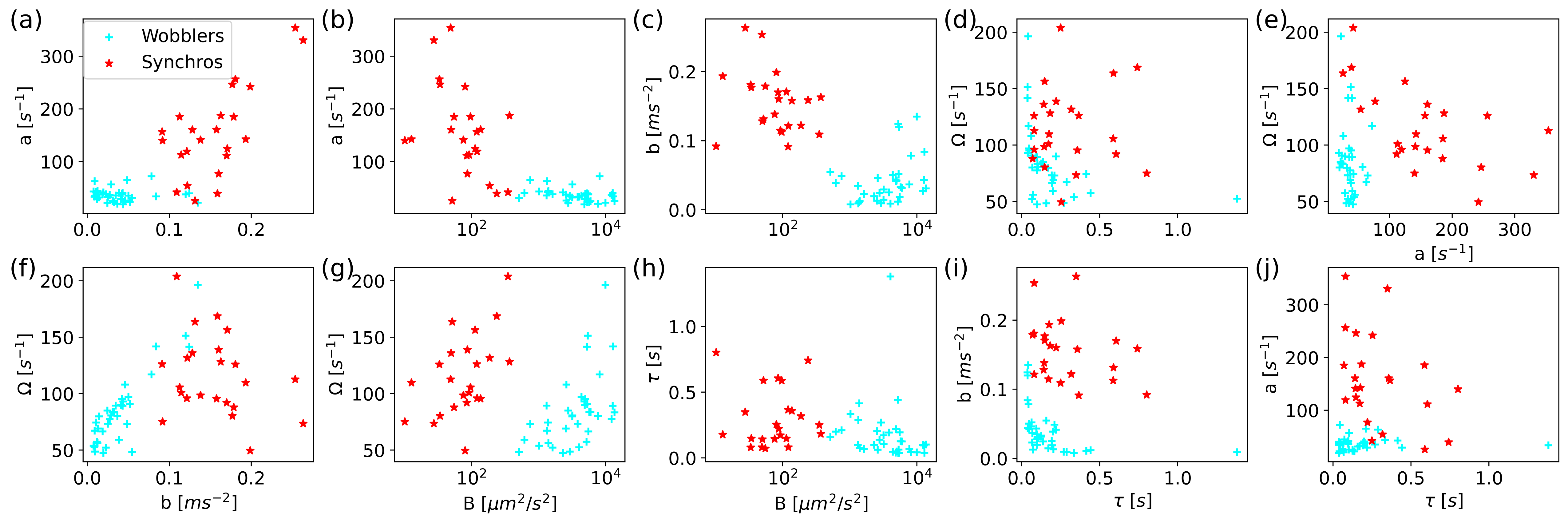}
	\caption{
		Scatter correlation plots of   individual CR cell parameters, consisting of  the friction kernel 
		parameters ($a$, $b$, $\tau$, $\Omega$) defined in eq. \eqref{eq_kernel_osc_def}  and the mean-squared velocity $B$. 
		All parameters except $B$ are presented on a linear scale. 
		Synchros are shown in red  and wobblers in cyan according to our cluster analysis, 
		which perfectly matches a categorization based on the visual analysis of flagella motion.}
	\label{fig_5}
\end{figure*}

{\it Clustering of single-cell parameters:}
The GLE \eqref{eq_gle} in conjunction with the random force strength $B$ defined in eq. \eqref{eq_fdt_neq}
and  the effective friction kernel eq. \eqref{eq_kernel_osc_def} 
has five parameters. This gives rise to 10  distinct two-dimensional projections, which 
are shown in Fig. \ref{fig_5}. Each data point corresponds to a single cell.
The parameters exhibit  substantial spread among individual cells, but an unambiguous  separation between 
wobbler and synchros, here colored in blue and red, is not obvious. 
As can already be seen in Figs. \ref{fig_3}c,f, the friction amplitudes $a$ and $b$ are larger for the synchros,
while the mean squared velocity $B$ is larger for wobblers, which leads to a separation of the two populations
in Figs. \ref{fig_5}b,c.
Each flagellar beating cycle leads to a net forward cell motion, 
which is reflected by the positive correlation between memory 
oscillation frequency $\Omega$ and mean-squared velocity $B$ in Fig. \ref{fig_5}g for each cell type.

We  perform an unbiased cluster analysis using the X-means algorithm \cite{pelleg2000xmeans}, which is a general version of the k-means algorithm \cite{bock2007clustering} that self-consistently
 determines the  optimal number of clusters (details are given in Methods \ref{sec_cluster_analysis}).
Applying our unbiased cluster analysis to the five dimensional parameter space, 
we obtain  two distinct groups, which perfectly coincide with the assignment into wobblers and synchros 
 from visual  analysis of the flagellar motion in the video data \cite{mondal_strong_2021}.
This means that we can   classify the cells by just knowing their center of mass trajectories with 
 an accuracy of $100\,\%$. In comparison, a  cluster analysis solely based on the mean-squared velocity $B$ 
 leads  to an accuracy of only $69\,\%$, while a cluster analysis using the first two PCA components leads to an accuracy of $90\,\%$ 
 (see SI Sec. VIII). Using only the four friction kernel parameters for a cluster analysis 
 without the mean squared velocity $B$ still reaches an accuracy of $92\,\%$.
This clearly demonstrates that the GLE model, which parameterizes cell motion based on 
friction kernel parameters in  eq. \eqref{eq_kernel_osc_def} together with the mean squared velocity $B$,
allows for accurate  classification of cells based only on their motion.

	\section{Discussion and Conclusions}

We demonstrate that the rather complex motion of individual CR algae   can be accurately 
described by the GLE  \eqref{eq_gle} and extract  all  GLE model parameters  for individual algae from 
their cell-center trajectories in a data-driven manner.
Based on the extracted GLE parameters, we detect  two distinct algae classes by  an unbiased cluster analysis,
this unsupervised clustering result is confirmed by comparison with a categorization based on visual inspection of the flagella beating patterns.
Our method is applicable to any kind of cells and even higher organisms. 
Since the GLE \eqref{eq_gle} is not restricted to positional degrees of freedom, 
our approach can be applied to any observable, for instance cell extension or deformation.
As  the only input  needed are trajectories, our method requires minimal interaction with the 
organisms  and can be easily used as a stand-alone tool or to improve existing machine-learning algorithms for cell classification. 

 Moreover, our approach allows  for a mechanistic interpretation of cell-motion characteristics.
In fact, a friction kernel model that is very similar to  eq. \eqref{eq_kernel_osc_def} and  describes the data  equally well can be derived from the equation of motion of two elastically
 coupled objects (see Methods \ref{sec_markovian_embedding} and SI Sec. IX for details).
Without further experimental input our approach does not reveal what these objects are,
so we can only speculate that the elastic coupling between the cell body and the flagella,
which presumably involves the connection between 
the flagellar basal bodies by the distal striated fiber (schematically shown in Fig. \ref{fig_1}c),
causes the slowly decaying oscillations in the memory kernel.
This seems in line with 
previous models for CR algae motion and flagella synchronization \cite{friedrich2012flagellar,quaranta2015hydrodynamics_vs_intra}.
Clearly, more experiments that resolve the relative motion of the cell center and the flagella
are needed to resolve these issues.

Our extraction of GLE parameters from cell trajectories yields a single effective memory kernel 
and does not allow to detect the non-equilibrium character of cell motion, 
in agreement with recent general arguments \cite{roland_neq_2023b}.
Conversely, based on an explicit non-equilibrium model for cell motion, it is rather
straightforward to derive the GLE  \eqref{eq_gle} and the functional form of the extracted kernel eq. \eqref{eq_kernel_osc_def}, as shown in  SI Sec. X.
A similar GLE model can also be derived for motion in a confining potential to describe confined neurons \cite{zeng2017neurons_classification} or bacteria moving in mucus \cite{ribbeck2021mucus}
(see SI Sec. III for details).

As a final note, we mention that the angular orientation of CR algae 
can be accurately extracted from  their cell-center trajectory, 
as shown in SI Sec. XI.
Thus, the  orientational cell dynamics is included in our GLE model.

In summary, our approach allows for cell classification by positional cell-center
trajectories or any other kind of time series data and at the same time for interpretation of the motion pattern
in terms of intracellular interactions. 
We anticipate numerous applications in biology and medicine 
that require the label-free distinction of individual cells and organisms.

	\section*{Acknowledgement}
	We acknowledge  funding by the Deutsche Forschungsgemeinschaft (DFG) through grant CRC 1449 “Dynamic Hydrogels at Biointerfaces”, Project ID 431232613, Project A03,  by the ERC Advanced Grant NoMaMemo No. 835117
	and by the Infosys Foundation. 
	D.M. and P.S. acknowledge support from the DBT/Wellcome Trust India Alliance Fellowship [grant number IA/I/16/1/502356] awarded to P.S.

	\bibliography{cell_classification.bib}
	
	\section{Methods}
	
	\subsection{Cell growth and sample preparation}   \label{sec_cell_preparation}
Wild-type CR cell cultures (strain, CC-1690) are grown in Tris-Acetate-Phosphate (TAP+P) medium by alternating light:dark (12:12 hr) cycles for three days. We collect the cell suspension in their actively growing phase 
(between 3rd and 6th day of culture) 2-3 hours after the beginning of the  light cycle and re-suspend it in fresh TAP+P medium. After 30 to 40 minutes of equilibriation to recover from the mechanosensitive shock during re-suspension \cite{Wakabayashi2020mechanosensitive}, we inject the cells inside a rectangular quasi-2D microfluidic chamber of height $ 10\, \mu m$ and area $18\, mm \times 6\, mm$. This chamber is assembled by using a glass slide and coverslip sandwiched with a $ 10\, \mu m$ double-sided tape (Nitto Denko corporation) as spacer. The glass surfaces are pre-cleaned and coated with 
a polyacrylamide brush to suppress non-specific adhesion of cell body and flagella \cite{mondal_2020_cilia}.  
The chamber height is determined as  $ 10.88 \pm 0.68 \, \mu m$ across different samples \cite{mondal_strong_2021}.
	
	\subsection{Recording of trajectories}   \label{sec_recording}
The cells in the chamber are placed  under red light illumination ($> 610 \, nm$) to prevent phototaxis \cite{Goldstein2015Rev} and flagellar adhesion \cite{Baumchen2018NatPhys} of CR \cite{mondal_strong_2021}. 
We use high-speed video microscopy (Olympus IX83/IX73) at 500 frames per second with a 40X phase-contrast objective (Olympus, 0.65 NA, Plan N, PH2) connected to a  metal oxide semiconductor (CMOS) camera (Phantom Miro C110, Vision Research, pixel size = $ 5.6 \, \mu m$) for imaging the mid-plane between the confining  glass plates. 
This setup enables us to simultaneously image  cell position and flagellar shape.
To capture  very  long trajectories to probe the long-time diffusive behavior  in SI Sec. IV, 
we use a 10X bright-field objective (Olympus, 0.25 NA, PlanC N) connected to a high-speed CMOS camera of higher pixel length (pco.1200hs, pixel size = $ 12 \, \mu m$) at 50 frames per second. 
We determine  cell trajectories by binarizing the image sequences with appropriate threshold parameters and tracking their 
centers using standard MATLAB routines \cite{Matlabtrack}.

	\subsection{Velocity autocorrelation function}
	\label{sec_mapping_derivation}
	Fourier transformation of eq. \eqref{eq_gle} and eq. \eqref{eq_fdt_neq} leads to
	\begin{equation}
		\label{eq_ft_gle_neq}
		\tilde{v}(\omega)=\frac{\tilde{F}_R(\omega)}{\tilde{\Gamma}^+_v(\omega) +i\omega}
	\end{equation}
with the  single-sided Fourier transform defined as 
$\tilde{\Gamma}^+_v(\omega) = \int_{0}^{\infty} e^{-i\omega t} {\Gamma}_v(t) dt$ and 
	\begin{equation}
		\label{eq_ft_force_corr}
		\langle  \tilde{F}_R(\omega) \tilde{F}_R(\omega') \rangle = 2\pi B \delta(\omega+\omega')\tilde{\Gamma}_R(\omega')\,.
	\end{equation}
From the Fourier transform of the VACF
	\begin{equation}
		\begin{split}
		\label{eq_ft_vacf}
		&\tilde{C}_{vv}(\omega) = \int_{-\infty}^{\infty} dt e^{-i\omega t}\langle v(0)v(t)  \rangle \\
		&= \int_{-\infty}^{\infty}e^{-i\omega t} dt \int_{\infty}^{\infty}e^{i\omega t} \frac{d\omega}{2\pi} \int_{-\infty}^{\infty} \frac{d\omega}{2\pi} 
		\langle \tilde{v}(\omega) \tilde{v}(\omega')  \rangle
		\end{split}
	\end{equation}
we obtain by inserting eqs. \eqref{eq_ft_gle_neq} and \eqref{eq_ft_force_corr} 
	\begin{equation}
		\label{eq_ft_vacf_final}
		\tilde{C}_{vv}(\omega) = \frac{B \tilde{\Gamma}_R(\omega)}{(\tilde{\Gamma}_v^+(\omega) +i\omega) (\tilde{\Gamma}_v^+(-\omega) -i\omega)} \,.
	\end{equation}
	Equating the non-equilibrium  and the surrogate VACF in eqs. \eqref{eq_condition_eq_neq}
and \eqref{eq_surr} leads to 
	\begin{equation}
		\label{eq_condition_eq_neq2}
		\frac{\tilde{\Gamma}_R(\omega)}{(\tilde{\Gamma}_v^+(\omega) +i\omega) (\tilde{\Gamma}_v^+(-\omega) -i\omega)} =
		\frac{\tilde{\Gamma}(\omega)}{\mid\tilde{\Gamma}_+(\omega) +i\omega\mid^2} \,.
	\end{equation}
In SI Sec. II we show that for every correlation function $C_{vv}(t)$ we can determine
 a unique $\Gamma(t)$.

	\subsection{Memory kernel extraction}
	\label{sec_kernel_extraction}
Multiplying the GLE eq. \eqref{eq_gle} by $\dot{x}(t_0)$, averaging over the random force
 and integrating from $t_0$ to $t$ leads to
	\begin{equation}
		\label{eq_volterra}
		(C_{vv}(t)-C_{vv}(0)) = -\int_{0}^{t} C_{vv}(s) G(t-s) ds \,,
	\end{equation}
	where we used that  $\langle \dot{x}(t_0) F_R(t)\rangle =0 $ \cite{mori_transport_1965,zwanzig1961memory,cihan2022_hybrid_gle},
set  $t_0=0$  and introduced the integral kernel 
	\begin{equation}
		\label{eq_G_def}
		G(t) = \int_{0}^{t} \Gamma (s) ds \,.
	\end{equation}
In order to invert eq. \eqref{eq_volterra}, we discretize it.
Since $C_{vv}(t)$  is even while $G(t)$  is odd, we
discretize  $G(t)$ on half steps and $C_{vv}(t)$ on full steps and obtain \cite{mitterwallner_non-markovian_2020} 
	\begin{equation}
		\label{eq_G_half_step}
		G_{i+1/2} = \frac{2(C_{vv}^0 -C_{vv}^{i+1})}{\Delta(C_{vv}^1 + C_{vv}^0)} - \sum_{j=1}^{i} G_{i-j+1/2}\frac{C_{vv}^{j+1}+C_{vv}^j}{C_{vv}^1+C_{vv}^0} \,.
	\end{equation}
The kernel $\Gamma_i$ is  obtained by  the discrete derivative $\Gamma_i=\frac{G_{i+1/2} + G_{i-1/2}}{\Delta}$
with the initial value  $\Gamma_0 = 2G_{1/2}/\Delta$.	
	
	\subsection{Two-point velocity distribution}
	\label{sec_greens_func}
A stationary Gaussian process is completely described by its two-point probability distribution as shown in SI Sec. III.
 Here we show that the joint and conditional velocity  distributions only depend on the VACF. 
 The joint probability to observe $v_2$ at time $t_2$ and $v_1$ at time $t_1$ can be written in 
 terms of the velocity vector $\vec{v}=(v_1,v_2)^T$ as 
	\begin{equation}
		\label{eq_joint_prob}
		p(v_2,t_2;v_1,t_1)=\frac{\exp(-\vec{v}^T\Sigma^{-1}(t_2-t_1)\vec{v}/2)}{\sqrt{2\pi\mid\Sigma(t_2-t_1)\mid}}
	\end{equation}
	with ${\mid\Sigma(t_2-t_1)\mid}$ denoting the determinant of the covariance matrix
	\begin{equation}
		\begin{split}
		\label{eq_covariance_matrix_vel}
		\Sigma &= \begin{pmatrix}
			\langle v(t_1)v(t_1) \rangle & \langle v(t_1)v(t_2) \rangle \\
			\langle v(t_1)v(t_2) \rangle & \langle v(t_2)v(t_2) \rangle
		\end{pmatrix}\\
		&= \begin{pmatrix}
			C_{vv}(0) & C_{vv}(t_2-t_1)\\
			C_{vv}(t_2-t_1) & C_{vv}(0)
		\end{pmatrix} \,.
		\end{split}
	\end{equation}
	Using the normally distributed velocities
	\begin{equation}
		\label{eq_vel_distr_gauss}
		p(v) = \frac{\exp(\frac{-v^2}{2C_{vv}(0)})}{\sqrt{2\pi C_{vv}(0)}} \,,
	\end{equation}
	we obtain  the  conditional probability that $v(t_2)=v_2$ given $v(t_1)=v_1$, as
	\begin{equation}
		\begin{split}
		\label{eq_conditional_prob_vel}
		&p(v_2,t_2\mid v_1,t_1) = \frac{p(v_2,t_2;v_1,t_1)}{p(v_1, t_1)}\\
		&= \frac{\exp\left( -\frac{[v_2-v_1(C_{vv}(t_2-t_1))/C_{vv}(0)]^2}{(C_{vv}(0)-C_{vv}^2(t_2-t_1)/C_{vv}(0))} \right)}{\sqrt{2\pi (C_{vv}(0)-C_{vv}^2(t_2-t_1)/C_{vv}(0))}} \,.
		\end{split} 
	\end{equation}
Thus, the joint and conditional distributions only depend on the VACF $C_{vv}(t)$.

	\subsection{Fitting of friction kernel}
	\label{sec_fitting_kernel_directly}

 The extracted friction kernel of each individual cell is  fitted to eq. \eqref{eq_kernel_osc_def} by  least-square minimization 
 (using the curve fit function of python's \textit{scipy}) of the first $0.2\,s$ of the data.  
 The $\delta$-peak in eq. \eqref{eq_kernel_osc_def} leads to  the initial kernel value  ${\Gamma_0 = 2a/\Delta +b}$ where $\Delta$ denotes the discretization time, see Sec. \ref{sec_kernel_extraction}.
Since single cells exhibit large variations of  the friction kernels, we constrain $a$ and $b$ in
 eq. \eqref{eq_kernel_osc_def} to be between $0.1\,\%$ and $99.9\,\%$ of  $\Gamma_0$. 
 The decay time $\tau$ is constrained to be between $0.05\,s$ and $3\,s$ 
 and the frequency $\Omega$ is constrained between $20\,s^{-1}$ and $250\,s^{-1}$.

	\subsection{Discretized VACF including localization noise}\label{sec_fitting}
	In order to test whether the cell motion  is actually described by the friction kernel eq. \eqref{eq_kernel_osc_def} and not originating from the experimental finite time step or noise, we fit the experimental VACF of individual cells with 
an analytical model that accounts for finite time discretization and noise  \cite{mitterwallner_non-markovian_2020}. 
Here we explain the fitting procedure, the analytical expression for the MSD using a friction kernel in the form
of eq. \eqref{eq_kernel_osc_def} is derived in SI Sec. VI.

We denote discrete values of a function $f(t)$ as
 $f(i \Delta)=f_i=f^i$ and the discretization time step as $\Delta$.
 After smoothing the data by averaging over consecutive positions to reduce the localization noise, 
 as discussed in detail in SI Sec. XII, the velocities at half time steps follow as 
	\begin{equation}
		\label{eq_central_diff_vel}
		v_{i+\frac{1}{2}} = \frac{x_{i+1} - x_{i}}{\Delta}\,.
	\end{equation}
	From the velocities the VACF defined by eq. \eqref{eq_def_cvv}
	is calculated  according to
	\begin{equation}
		\label{eq_vacf_from_data}
		C_{vv}^i=\frac{1}{N-i}\sum_{j=0}^{N-i}v_{j+\frac{1}{2}} v_{j+i+\frac{1}{2}}\,,
	\end{equation}
	with $N$ being the the  number of  trajectory steps.
In order to account for localization noise, we assume Gaussian uncorrelated  noise of width $\sigma_{loc}$ at every time step, which gives the noisy  MSD as \cite{mitterwallner_non-markovian_2020}
	\begin{equation}
		\label{eq_msd_noise}
		C_{\textrm{MSD}}^{\rm{noise}}(t)=C_{\textrm{MSD}}^{\rm{theo}}(t)+2(1-\delta_{t0})\sigma_{\textrm{loc}}^2 \,,
	\end{equation}
where $C_{\textrm{MSD}}^{\rm{theo}}(t)$ is the theoretical expression for the model MSD given in 
SI Sec. VI, eq. (S51),  
and $\delta_{t0}$ is the Kronecker delta reflecting the uncorrelated nature of the localization noise.
	Since the observed trajectories are sampled with a finite time step $\Delta$, we discretize  the relation
	\begin{equation}
		C_{vv}(t) = \frac{1}{2}\frac{d^2}{dt^2}C_{\textrm{MSD}}(t)\,,
		\label{eq_cvv_from_msd}
	\end{equation}
	which leads to
	\begin{equation}
		\hspace{-7mm}
		\label{eq_vacf_fit}
		\begin{split}
			&C_{vv}^{\rm{fit}}(i \Delta)=\\
			&\frac{C_{\textrm{MSD}}^{\rm{noise}}((i+1)\Delta)-2C_{\textrm{MSD}}^{\rm{noise}}(i \Delta)+C_{\textrm{MSD}}^{\rm{noise}}((i-1)\Delta)}{2\Delta^2} \hspace{1mm}.
		\end{split}
	\end{equation}
Finally, fits are performed by minimizing the cost function
	\begin{equation}
		\label{eq_cost_function}
		E_{\textrm{cost}}=\sum_{i=0}^{n} (C_{vv}^{\rm{exp}}(i \Delta)-C_{vv}^{\rm{fit}}(i \Delta))^2
	\end{equation}
	with scipy's least squares function in python and using eq. \eqref{eq_vacf_fit} to determine $C_{vv}^{\rm{fit}}(t)$ at discrete time points. As the MSD and VACF follow from the GLE 
\eqref{eq_gle} and the friction kernel eq. \eqref{eq_kernel_osc_def}, the parameters to optimize are the kernel parameters $a$, $b$, $\tau$, $\Omega$, the mean squared velocity $B$ and the localization noise width $\sigma_{\rm{loc}}$. 
Resulting fits  are shown in Figs. \ref{fig_4}b,e, where the data is fitted up to $0.2\,s$ in order to 
disregard the noisy part of the VACF (see SI Sec. VII for details).	
	
	\subsection{Cluster analysis}
	\label{sec_cluster_analysis}
The friction kernel in eq. \eqref{eq_kernel_osc_def} contains four parameters, together with the mean-squared velocity $B$ each individual cell is characterized by five parameters.
We perform an X-means cluster analysis \cite{pelleg2000xmeans}, which is a generalization of the k-means algorithm \cite{bock2007clustering}.
The k-means algorithm assigns unlabeled data to a predetermined number of $k$ clusters by minimizing distances to the cluster centers.
In  the X-means algorithm  the number of clusters is not predetermined, we allow  cluster numbers from $2$ to $20$. 
The algorithm starts with the minimal number of clusters and  finds the cluster centers
using k-means. It then  splits every cluster into two subclusters whose centers are again determined by k-means. 
New subclusters are accepted if they improve the clustering quality  accounting for the increased number of parameters. 
For this we use the minimal noiseless description length (MNDL) criterion \cite{beheshti2005new,shahbaba2012improvingxmeans}.
We use an implementation of the X-means algorithm in python \cite{Novikov2019_python} and  use individual cell parameters as initial cluster centers \cite{celebi2013comparative}.
 The X-means algorithm can converge to different final  results depending on the initial cluster-centers. 
 Thus, we use all $59\cdot58/2$ possible combinations of initial cluster centers and use the result that occurs most often.
We rescale each parameter by the median of its distribution.

	\subsection{Markovian embedding}
	\label{sec_markovian_embedding}
A similar kernel to eq. \eqref{eq_kernel_osc_def}, namely
	\begin{equation}
		\label{eq_kern_markov_embedding}
		\Gamma(t)=2a\delta(t) + be^{-t/\tau}\left(\cos(\Omega t) + \frac{1}{\tau\Omega}\sin(\Omega t)\right) \,,
	\end{equation}
can be derived from a system of harmonically coupled particles.
In fact, eq.  \eqref{eq_kern_markov_embedding}
 becomes equivalent to eq. \eqref{eq_kernel_osc_def} if the 
   oscillation period $1/\Omega$ is much smaller than the decay time $\tau$,
   which is the case for the extracted algae kernels.
	The Hamiltonian describing the coupled particles takes the form
	\begin{equation}
		\label{eq_hamiltonian_osc}
		H = \frac{m}{2} v^2 + \frac{m_y}{2} v_y^2 + \frac{K}{2}(x - y)^2 \,,
	\end{equation}
where $m_i$ are the  masses of the two particles and $K = b m$ is the harmonic coupling strength.
In the presence of friction, quantified by  friction coefficients $\gamma_i$ for each component,
and coupling the particles to a heat bath at temperature $T$, 
the  coupled equations of motion are  given by

	\begin{equation}
		\begin{aligned}
		\label{eq_markov_embedding}
		\dot{x}(t) &= v(t) \\
		m \dot{v}(t) &= -\gamma_x v + b m (y(t) - x(t)) + F_{Rx}(t) \\
		\dot{y}(t) &= v_y(t) \\
		m_y \dot{v}_y(t) &= -\gamma_y v_y + b m (x(t) - y(t)) + F_{Ry}(t) \,,
		\end{aligned}
	\end{equation}
	where $F_{Rx}(t)$ and $F_{Ry}(t)$ are random forces with zero mean and second moment $\langle F_{Ri}(0)F_{Rj}(t)\rangle = \delta_{ij} 2\gamma_i k_BT \delta(t)$. 
In SI Sec. XIII it is shown that the  coupled equations of motion \eqref{eq_markov_embedding}
are equivalent to a GLE in the form of eq. \eqref{eq_gle} for the coordinate $x(t)$ with a memory kernel $\Gamma(t)$ given by eq. \eqref{eq_kern_markov_embedding}\cite{brunig2022time}. The friction of the first particle leads to the $\delta$-contribution of $\Gamma(t)$
and the harmonic coupling to the second  particle leads to the oscillating exponentially decaying contribution. 
The parameters of eq. \eqref{eq_markov_embedding} translate into the parameters of eq. \eqref{eq_kern_markov_embedding} as 
	\begin{eqnarray}
		\label{eq_embedding_params}
		a &=& \frac{\gamma_x}{m} \nonumber\\
		\tau &=& 2\frac{m_y}{\gamma_y} \\
		\Omega &=& \sqrt{\frac{b m}{m_y} - \frac{1}{\tau^2}} \nonumber \,.
	\end{eqnarray}

\end{document}


\title[ ]{Supplementary information:
		Data-driven classification of single cells by their non-Markovian motion}
	\maketitle
	
	\tableofcontents
	
	\newpage
	
	\section{Video captions}\label{sec_video_caption}
	\textbf{Video 1.: A representative synchro.}
	High-speed video microscopy at 500 frames per second obtained by phase-contrast imaging of a synchro CR cell showing the planar and synchronous breaststroke motion of the flagella. Between t $ \sim 330 - 390 \, ms $ the synchronous beat of the flagella exhibits a phase slip, meaning the synchronicity of the flagella is disturbed in that short time interval. 
	\\
	\textbf{Video 2.: A representative wobbler.}
	High-speed video microscopy at 500 frames per second obtained by phase-contrast imaging of a CR cell which paddles the flagella in an asynchronous and irregular manner resulting in the wobbling motion of the cell body.
	
	\section{Effective kernel follows uniquely  from correlation functions}\label{sec_mapping_SI}
	We can write the GLE with an arbitrary potential $U(x(t))$ as
	\begin{equation}
		\ddot{x}(t) = -\nabla U(x(t)) -\int_{t_0}^{t} \Gamma_v(t-t')\dot{x}(t') dt' + F_R(t)\,.
		\label{eq_gle_harm}
	\end{equation}
	Multiplying eq. \eqref{eq_gle_harm}
	by $v(t_0)$, inserting $\Gamma(\vert t\vert)=\Gamma_R(t)=\Gamma_v(\vert t \vert)$ and averaging the entire equation leads to the Volterra equation
	\begin{equation}
		\label{eq_volterra_pot}
		\dfrac{d}{dt}C_{vv}(t) = - C_{v\nabla U}(t) -\int_{0}^{t}\Gamma(t-s)C_{vv}(s)ds  \,,
	\end{equation}
	using that the random force is not correlated with the initial velocity at the projection time, i.e. $\langle v(t_0)F_R(t) \rangle = 0$ and setting $t_0=0$.
	Volterra equations are well-behaved  and for given  functions $\Gamma(t)$ and $C_{v\nabla U}(t)$ one can find a solution of eq. \eqref{eq_volterra_pot} in terms of   $C_{vv}(t)$.
	Inversely,  for given correlation functions $C_{vv}(t)$ and $C_{v\nabla U}(t)$ one can solve eq. \eqref{eq_volterra_pot} in terms of the friction kernel $\Gamma(t)$ by  Laplace transformation.
	With the definition of the Laplace transform of a function $f(t)$ 
	\begin{equation}
		\hat{f}(q) = \int_0^{\infty}f(t) e^{-qt}dt,
		\label{eq_def_laplace_transform}
	\end{equation}
	the unique  solution for $\Gamma(t)$ in Laplace space is given by 
	is given by
	\begin{equation}
		\label{eq_solution_volterra_gamma_laplace}
		\hat{\Gamma}(q)=-\frac{1}{\hat{C}_{vv}(q)} \left( q\hat{C}_{vv}(q)-C_{vv}(0) + \hat{C}_{v\nabla U}(q) \right) \,.
	\end{equation}
	The existence of a unique friction kernel for any given input correlation functions $C_{vv}(t)$ and $C_{v\nabla U}(t)$
	assures that one can always find an effective friction kernel $\Gamma(t)$ that, when employed in the 
	GLE and using  $\Gamma(\vert t\vert) = \Gamma_R(t) = \Gamma_v(\vert t\vert)$,
	reproduces the two-point correlation functions.
	Thus, every non-equilibrium model described by  eq. (1) 
	with  $ \Gamma_R(t) \neq  \Gamma_v(\vert t\vert)$
	can be mapped on an effective model with $\Gamma(\vert t\vert) = \Gamma_R(t) = \Gamma_v(\vert t\vert)$, because  the Green's function, 
	which completely  describes a Gaussian process,
	is solely given in terms of  the positional  two-point correlation function, see 
	SI Sec. \ref{sec_ext_potential_SI}.

	\section{Green's function is given in terms of positional two-point correlation function}\label{sec_ext_potential_SI}
	In the following we  map a non-equilibrium process described by the GLE eq. \eqref{eq_gle_harm} with the random force fluctuations given by eq. (2) and a harmonic potential $U(x)=Kx^2/2$ on an effective model with an effective friction kernel $\Gamma(t)$ fulfilling $\langle F_R(0)F_R(t) \rangle = k_BT\Gamma(\vert t \vert)$ and an effective harmonic potential ${U_{\rm{eff}}(x)=K_{\rm{eff}}x^2/2}$.
	For this we show that  the Green's function for the non-equilibrium system eqs. \eqref{eq_gle_harm}, (2) is given in terms  of the positional autocorrelation function.
	
	We define the average over the random force by the path integral 
	\begin{equation}
		\langle X \rangle = \mathcal{N}\int\mathcal{D}\tilde{F}_R(\cdot) X \exp \left( -\frac{1}{4\pi B}\int_{-\infty}^{\infty}d\omega\int_{-\infty}^{\infty}d\omega'  \frac{\tilde{F}_R(\omega) \delta(\omega + \omega') \tilde{F}_R(\omega')}{\tilde{\Gamma}_R(\omega)} \right)\,.
	\end{equation}
	We also define the  generating  functional as
	\begin{align}
		\mathcal{Z}[\tilde{h}] =& \mathcal{N} \int\mathcal{D}\tilde{F}_R(\cdot) \exp \left( -\frac{1}{4\pi B}\int_{-\infty}^{\infty}d\omega \frac{\tilde{F}_R(\omega)  \tilde{F}_R(-\omega)}{\tilde{\Gamma}_R(\omega)} + \int_{-\infty}^{\infty}d\omega\tilde{h}(-\omega) \tilde{F}_R(\omega) \right) \nonumber \\
		=& \exp \left( 2\pi B \int_{-\infty}^{\infty}d\omega \tilde{h}(\omega)\tilde{h}(-\omega) \tilde{\Gamma}_R(\omega) \right) \label{eq_functional_rf_SI}
	\end{align}
	where $\tilde{h}(\omega)$ is a generating field. Equation \eqref{eq_functional_rf_SI}  has the functional derivatives
	\begin{align}
		&\dfrac{\delta\mathcal{Z}[\tilde{h}]}{\delta \tilde{h}(\omega)}\mid_{\tilde{h}=0} = \langle \tilde{F}_R(\omega) \rangle =0 \\
		&\dfrac{\delta^2\mathcal{Z}[\tilde{h}]}{\delta \tilde{h}(\omega) \delta \tilde{h}(\omega')}\mid_{\tilde{h}=0} = \langle \tilde{F}_R(\omega) \tilde{F}_R(\omega') \rangle = 2\pi B \delta(\omega+\omega') \tilde{\Gamma}_R(\omega)\,,
	\end{align}
	where $\tilde{\Gamma}_R(\omega) = \tilde{\Gamma}_R(-\omega)$.
	For  harmonic potential ${U(x)=Kx^2/2}$ we can  Fourier-transform  the GLE eq. \eqref{eq_gle_harm} 
	by which we obtain the form of eq. \eqref{eq_ft_gle} with the response function given by 
	\begin{equation}
		\tilde{\chi}(\omega) = \left(K -\omega^2 + i\omega\tilde{\Gamma}_v^+(\omega) \right)^{-1}\,.
		\label{eq_response_harm}
	\end{equation}
	The Green's function can be written as
	\begin{equation}
		G(x, t\mid x', t') = \frac{P(x, t; x', t')}{P(x', t')},
		\label{eq_def_green_ratio}
	\end{equation}
	which is the ratio of the joint two-point probability distribution to find $x$ at $t$ and $x'$ at $t'$ and
	the base probability to find $x'$ at $t'$.
	Those probabilities can  be expressed as
	\begin{equation}
		P(x', t') = \langle \delta(x'-x(t')) \rangle \, , \, P(x, t; x', t') = \langle \delta(x'-x(t')) \delta(x-x(t)) \rangle\,.
		\label{eq_delta_probs}
	\end{equation}
	From eq. \eqref{eq_ft_gle} we can write the solution of the GLE eq. \eqref{eq_gle_harm} 
	using the linear response function $\chi(t)$ as
	\begin{equation}
		x(t) = \int_{-\infty}^{\infty} \frac{d\omega}{2\pi}e^{i\omega t} \tilde{\chi}(\omega)\tilde{F}_R(\omega).
		\label{eq_xt_time_sol}
	\end{equation}
	Inserting  this result into the probability distributions  eq. \eqref{eq_delta_probs} we obtain
	\begin{equation}
		P(x', t') = \int_{-\infty}^\infty\frac{dp}{2\pi}e^{-ipx'} \Biggl\langle \exp\left( ip\int_{-\infty}^{\infty}\frac{d\omega}{2\pi}\tilde{\chi}(\omega)\tilde{F}_R(\omega)e^{i\omega t'} \right) \Biggl\rangle
		\label{eq_prob_base_expo_delta}
	\end{equation}
	and
	\begin{align}
		P(x, t; x', t') = \int_{-\infty}^\infty\frac{dp}{2\pi} \int_{-\infty}^\infty\frac{dq}{2\pi} e^{-iqx - ipx'} &\Biggl\langle \exp\Biggl( iq\int_{-\infty}^{\infty}\frac{d\omega}{2\pi}\tilde{\chi}(\omega)\tilde{F}_R(\omega)e^{i\omega t} \nonumber\\
		&+ip\int_{-\infty}^{\infty}\frac{d\omega}{2\pi}\tilde{\chi}(\omega)\tilde{F}_R(\omega)e^{i\omega t'} \Biggl) \Biggl\rangle.
		\label{eq_prob_joint_expo_delta}
	\end{align}
	By comparing eqs. \eqref{eq_prob_base_expo_delta}, \eqref{eq_prob_joint_expo_delta} to eq. \eqref{eq_functional_rf_SI} we find the generating field $\tilde{h}(\omega)$ for the base probability $P(x', t')$
	\begin{equation}
		\tilde{h}_b(-\omega) = \frac{ip}{2\pi} \tilde{\chi}(\omega)e^{i\omega t'}
	\end{equation}
	and for the joint probability $P(x, t; x', t')$ we find
	\begin{equation}
		\tilde{h}_j(-\omega) = \frac{i}{2\pi} \tilde{\chi}(\omega)\left( qe^{i\omega t} + pe^{i\omega t'} \right)\,,
	\end{equation}
	where the subscript $b$ stands for the base probability and $j$ for the joint probability.
	Using  that the position correlation function can be written as
	\begin{equation}
		C_{xx}(t-t') = B \int_{-\infty}^{\infty}\frac{d\omega}{2\pi}\tilde{\chi}(\omega)\tilde{\Gamma}_R(\omega)\tilde{\chi}(-\omega)e^{i\omega(t-t')} \label{eq_int_cxx},
	\end{equation}
	we arrive at the Gaussian form
	\begin{align}
		P(x', t') =& \int_{-\infty}^{\infty}\frac{dp}{2\pi}\exp \left( -C_{xx}(0)p^2 -ipx' \right)\nonumber\\
		=& \sqrt{\frac{1}{2\pi C_{xx}(0)}}\exp \left( -\frac{x'^2}{2C_{xx}(0)} \right) \label{eq_prob_base_gaussian}
	\end{align}
	and
	\begin{align}
		&P(x, t; x', 0) = \int_{-\infty}^{\infty}\frac{dp}{2\pi} \int_{-\infty}^{\infty}\frac{dq}{2\pi} \exp \left( -\frac{q^2+p^2}{2}C_{xx}(0) - qp C_{xx}(t) - iqx -ipx' \right) \nonumber\\
		=& 2\sqrt{\frac{1}{2(C_{xx}(0)^2-C_{xx}(t)^2)}} \exp \left( -\frac{(x'+x)^2}{4(C_{xx}(0)+C_{xx}(t))} - \frac{(x'-x)^2}{4(C_{xx}(0)-C_{xx}(t))}\right) \label{eq_prob_joint_gaussian}\,.
	\end{align}
	For simplicity we set $t'=0$ as the position correlation function only depends on the time difference $t-t'$.
	Inserting eqs. \eqref{eq_prob_base_gaussian} \eqref{eq_prob_joint_gaussian} into eq. \eqref{eq_def_green_ratio},
	we finally obtain the position Green's function as a Gaussian
	\begin{equation}
		G(x,t \mid x',0) = \frac{\exp(-\frac{(x-\mu)^2}{2\sigma^2})}{\sqrt{2\pi\sigma^2}}
		\label{eq_greens_function_neq}
	\end{equation}
	with mean and standard deviation
	\begin{equation}
		\mu = \mu(x',t) = \frac{C_{xx}(t)}{C_{xx}(0)}x' \,,\, \sigma^2= \sigma^2(t) =  C_{xx}(0) \left(1 -\frac{C_{xx}(t)^2}{C_{xx}(0)^2}\right) \label{eq_mean_std_green}\,.
	\end{equation}
	Thus  the Green's function is  entirely described in terms of the two-point positional correlation function $C_{xx}(t)$, note 
	that $C_{vv}(t) = -\dfrac{d^2}{dt^2}C_{xx}(t)$. 
	
	We next calculate the 
	effective harmonic  potential strength $K_{\rm{eff}}$.
	The condition to find an effective description of the GLE eq. \eqref{eq_gle_harm} arises from setting eq. \eqref{eq_int_cxx} equal to its effective version
	\begin{equation}
		\tilde{\chi}(\omega)\tilde{\Gamma}_R(\omega)\tilde{\chi}(-\omega)=\tilde{\chi}_{\rm{eff}}(\omega)\tilde{\Gamma}(\omega)\tilde{\chi}_{\rm{eff}}(-\omega) \label{eq_eq_noneq_cond_cxx}\,,
	\end{equation}
	which assures that the effective model has  the same Green's function as the non-equilibrium model, as  shown by eqs. \eqref{eq_greens_function_neq} and \eqref{eq_mean_std_green}. Here, $\tilde{\chi}_{\rm{eff}}(\omega)$ is the effective response function and $\Gamma(t)$ is the effective kernel.
	Inserting eq. \eqref{eq_response_harm} into eq. \eqref{eq_eq_noneq_cond_cxx}  leads to
	\begin{equation}
		\hspace{-11mm}
		\frac{\tilde{\Gamma}_R(\omega)}{(K-\omega^2+i\omega\tilde{\Gamma}_v^+(\omega))(K-\omega^2-i\omega\tilde{\Gamma}_v^+(-\omega))} = \frac{\tilde{\Gamma}(\omega)}{(K_{\rm{eff}}-\omega^2+i\omega\tilde{\Gamma}_+(\omega))(K_{\rm{eff}}-\omega^2-i\omega\tilde{\Gamma}_+(-\omega))} \label{eq_eq_neq_cond_cxx_inserted_harm}\,.
	\end{equation}
	Defining the potential of mean force  for the  effective system according to 
	\begin{equation}
		U_{\rm{eff}}(x)=-B\ln(p(x)) = K_{\rm{eff}} x^2/2 ,
		\label{eq_free_energy}
	\end{equation}
	with $p(x)$ being the probability to observe $x$, and
	comparing eq. \eqref{eq_free_energy} to the long time limit of the non-equilibrium distribution eq. \eqref{eq_greens_function_neq}, namely
	\begin{equation}
		p(x) = \lim_{t\to\infty} G(x,t \mid x',0) = \frac{\exp(-\frac{x^2}{2 C_{xx}(0)})}{\sqrt{2\pi C_{xx}(0)}} \,,
		\label{eq_long_time_limit}
	\end{equation}
	we find the effective harmonic potential strength to be
	\begin{equation}
		K_{\rm{eff}} = \frac{B}{C_{xx}(0)} \,.
		\label{eq_effective_coupling_harm}
	\end{equation}
	Note that the effective  potential will in general deviate from the  potential in the non-equilibrium GLE, 
	i.e. $K_{\rm{eff}}\neq K$. The present calculation can also be done for a general multi-point distribution  \cite{roland_neq_2023b}.

	\section{Long-time MSDs }\label{sec_long_time_msd}
\begin{figure}
		\centering
		\includegraphics[width=\textwidth]{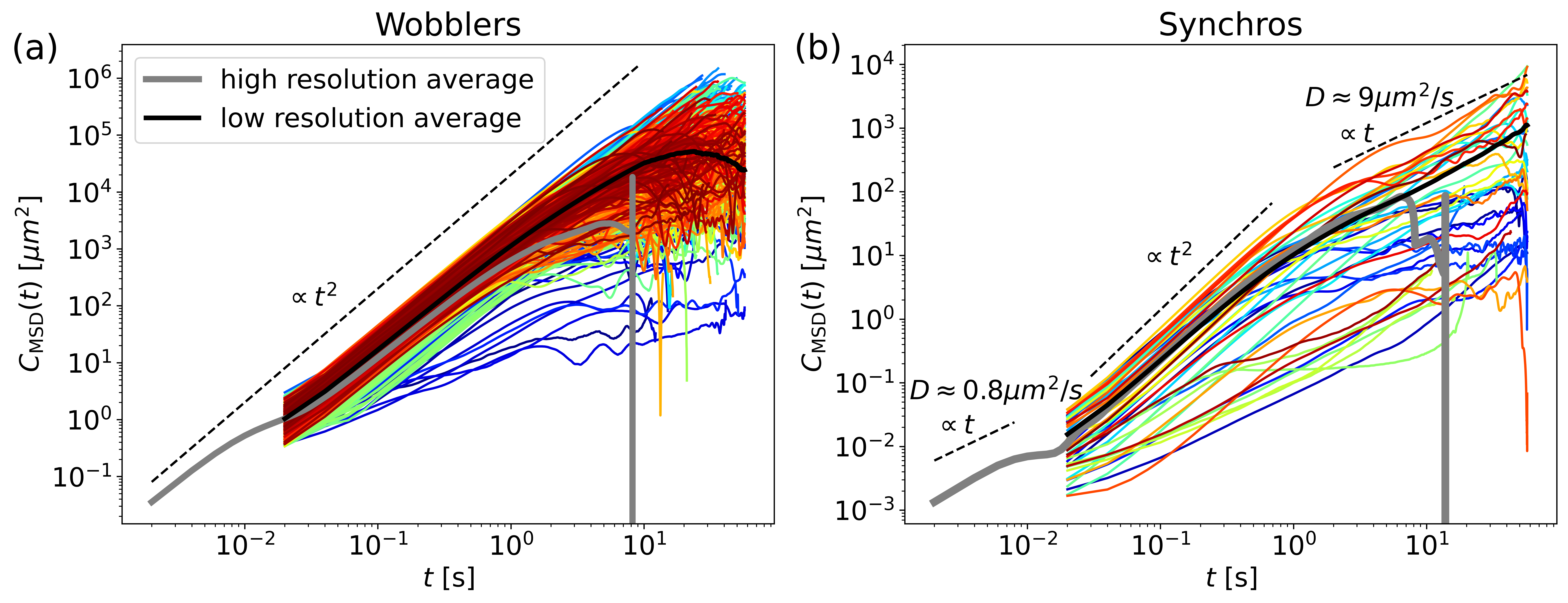}
		\caption{MSDs from low-resolution data with 10x magnification and a time step of $\Delta=0.02\,s$ of (a) wobblers  and (b) synchros according to an approximate classification scheme explained in the text. The
			average of the colored single-cell MSDs is shown in black,  the grey line represents the average MSD from Fig. 3 with higher temporal and spatial resolution of 40x and $\Delta=0.002\,s$.}
		\label{fig_putative_msd}
	\end{figure}

	The trajectories analyzed in the main text (examples shown in Figs. 2a,b) are recorded with a 40 fold magnification at 
	500 frames per second, which leads to trajectories of lengths up to $14\,s$ (Fig. 2h). Here, we analyze  a different data set to resolve the long time behavior of the MSD of the CR cells, which we obtain by using a lower magnification (tenfold) and longer time step of $0.02\,s$.
	A lower magnification results in a larger observation window, which in turn allows recording longer trajectories. The decreased localization precision and the longer time step do not allow to observe the details of the oscillatory features of the motion.
	Hence, we do not apply the cluster analysis relying on the extraction of the oscillating friction kernels, but rather use an approximate distinction by the mean squared velocities $B$.
	We assign cells with $B > 110\,\mu m^2/s^2$ to be putative wobblers and cells
	with $B < 35\,\mu m^2/s^2$ to be putative synchros.
	In this way, we avoid the assignment of cells   with intermediate mean squared velocities $B$,
	for which this approximate classification scheme is unreliable.
	For the slow cells we additionally exclude cells for which the MSD stays below $1\,\mu m^2$ during the entire trajectory. These cells are most likely stuck or have some defects prohibiting them to move. This distinction into putative wobblers and synchros is biased by our choice of the upper and lower bounds for $B$, which we choose according to the high resolution fastest synchro and slowest wobbler when using only every tenth step of the high resolution data in order to obtain comparable velocities with a time step of $0.02\,s$. Nevertheless, in Fig. \ref{fig_putative_msd} we demonstrate that  the present approximate classification
	leads to extended MSDs  (black lines) that match closely the MSDs shown in the man text based on the high-resolution videos
	(gray lines). The extended MSDs for  synchros exhibit a transition from the ballistic to the long time diffusive regime 
	at a time of about $t \approx 2$s, 
	whereas for the wobblers the ballistic regime extends over more than ten seconds.
	The prefactors of the  MSD in the diffusive regimes, i.e. the diffusivity $D$, 
	is for the synchros shown in Fig. \ref{fig_putative_msd}.
	
	\section{Relation between the oscillation periods of  the friction kernel and the VACF}\label{sec_frequency_vacf_kernel}
\begin{figure*}
		\centering
		\includegraphics[width=\textwidth]{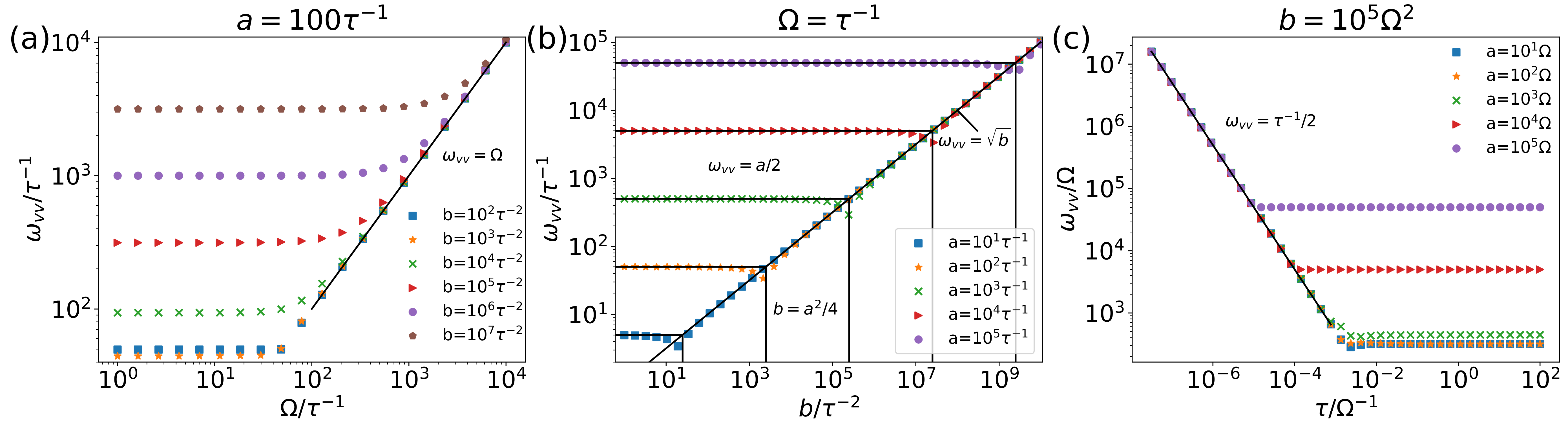}
		\caption{(a) The dependence of the VACF frequency $\omega_{vv}$ on the kernel frequency $\Omega$ of the kernel eq. (7) is shown for different kernel oscillation amplitudes $b$ and fixed $\delta$-peak amplitude $a=100\tau^{-1}$. The black line indicates the linear behavior $\omega_{vv}=\Omega$. (b) The dependence of the VACF frequency $\omega_{vv}$ on the kernel oscillation amplitude $b$ is shown for varying kernel $\delta$-peak amplitude $a$ and $\Omega=\tau^{-1}$, which represents the low kernel frequency plateau value shown in (a) for small $\Omega$. The horizontal lines represent $\omega_{vv}=a/2$ for the respective values of $a$, the diagonal line represents $\omega_{vv}=\sqrt{b}$ and the vertical lines represent the intersection of the other lines, which mark a minimum in the parameter $b$ positioned at $b=a^2/4$. (c) The dependence of the VACF frequency $\omega_{vv}$ on the kernel decay time $\tau$ shown for different $\delta$-peak amplitudes $a$ and fixed $b=10^5\Omega^2$. The regime shown for $\tau/\Omega^{-1}$ represents the low kernel frequency plateau value shown in (a) for small $\Omega\tau<100$. The black line indicates the scaling $\omega_{vv}=\tau^{-1}/2$.}
		\label{fig_frequencies}
	\end{figure*}

	In Fig. 4 in the main text one can note a subtle difference between the oscillation period of the friction kernel and the oscillation period of the VACF when comparing Fig. 4b to 4c or Fig. 4e to 4f. Since we derive an analytical expression for the VACF in eq. \eqref{eq_vacf_final_Res} for the friction kernel given in eq. (7), we can evaluate the dependence of the VACF oscillation frequency, denoted by $\omega_{vv}$, on the kernel frequency $\Omega$. From eq. \eqref{eq_vacf_final_Res} we know that there are three complex frequencies in the VACF that can lead to oscillations, where the frequency  $\omega_i$ with the smallest real part dictates $\omega_{vv}$. In Fig. \ref{fig_frequencies}a we show the dependence of $\omega_{vv}$ on $\Omega$ for different $b$ and find the VACF frequency $\omega_{vv}$ to be constant for small kernel frequencies $\Omega$ followed by a transition to the two frequencies being the same at a threshold value $\Omega_t=\omega_{vv}(\Omega\rightarrow 0)$, which is defined by the constant value $\omega_{vv}(\Omega\rightarrow 0)$ for small kernel frequencies intersecting with $\omega_{vv}=\Omega$. This threshold value depends on the kernel parameters $a$, $b$, $\tau$. It can be described by
	
	\begin{equation}
		\label{eq_threshold_freq} \Omega_t =
		\begin{cases}
			\ \sqrt{b} &, \text{for}\ b > \max (a^2/4, \tau^{-2}/4)\\
			\ a/2 &, \text{for}\ b<a^2/4 \land \tau>a^{-1}\\
			\tau^{-1}/2 &, \text{for}\ b<a^2/4 \land \tau<a^{-1}\,,
		\end{cases}
	\end{equation}
	which follows from Figs. \ref{fig_frequencies}a-c.
	
	Since the median kernel frequency of the CR cells is of the order ${\Omega\sim 100\,\tau^{-1}}$ and the oscillation amplitude is of the order $b\sim 10^4\tau^{-2}$, the data lies in the transition regime between the constant VACF frequency to the linear regime $\omega_{vv}=\Omega$, as can be seen in Fig. \ref{fig_frequencies}a. From Fig. \ref{fig_frequencies}a we know that the VACF frequency is always greater or equal to the kernel frequency, i.e. $\omega_{vv}\geq\Omega$. Moreover, the oscillation amplitudes of CR usually fulfill $b>a^2/4$ and the inverse decay time $\tau^{-1}$ is small compared to the other kernel parameters, such that $a>\tau^{-1}$ and $b>\tau^{-2}/4$ are always fulfilled. Therefore, the cells exhibit parameters for which $b > \max (a^2/4, \tau^{-2}/4)$ holds. Eq. \eqref{eq_threshold_freq} thus tells us that the threshold frequency is given by $\Omega_t=\sqrt{b}$. Combining this knowledge with $\omega_{vv}\geq\Omega$ leads to $\omega_{vv}\geq \sqrt{b}$. Since the CR data lies in the transition regime between $\omega_{vv}=\sqrt{b}$ and $\omega_{vv}=\Omega$, the VACF frequency $\omega_{vv}$ is strongly influenced by the oscillation amplitude $b$. In order to further estimate the exhibited parameter range, the reader is referred to Fig. 5 of the main text, which includes all parameters of all cells.
	
	The fact that the VACF frequency $\omega_{vv}$ increases with increasing kernel oscillation amplitude $b$ can be interpreted in the context of the Markovian embedding model in Sec. \ref{sec_markov_embedding}, where the coupling strength of the harmonically coupled parts is proportional to the oscillation amplitude $b$, as shown in eqs. \eqref{eq_markov_embedding_SI2}, \eqref{eq_markov_embedding_SI4}, which leads to faster oscillations for higher $b$. Since every oscillation leads to a net forward motion of the cell, it leads to fast net forward motion for cells if they exhibit a large kernel oscillation amplitude $b$ compared to the squared inverse decay time $\tau^{-2}$ and the squared $\delta$-amplitude $a^2$, which is the case in our data set (see Fig. 5).
	By realizing that the decay time $\tau$ is related to how long a cell 'remembers' its past trajectory, one can conclude that a fast decay of the oscillation would lead to a cell not being able to maintain its direction for a long time. Thus, it is useful to achieve fast net forward motion if the decay time is longer than the oscillation period, i.e. if $\tau\Omega>1$. By net forward motion we mean directed motion on relevant time scales, since of course in the long-time limit the cells exhibit diffusion as shown in Fig. \ref{fig_putative_msd}. Both criteria, the oscillation amplitude being large, i.e. $b>\max(\tau^{-2}, a^2)$ and the decay time being long compared to the oscillation period, i.e. $\tau\Omega>1$, are fulfilled by the CR cell parameters shown in Fig. 5, which means that CR cells exhibit a range of kernel parameters that is suited to achieve fast net forward motion.

	\section{Derivation of the analytical expression for the MSD}\label{sec_msd_ana}
	In Fourier space the GLE eq. (1) can be written as
	\begin{equation}
		\tilde{x}(\omega) = \tilde{\chi}(\omega)\tilde{F}_R(\omega) \,,
		\label{eq_ft_gle}
	\end{equation}
	where $\tilde{\chi}(\omega)$ is the Fourier transform of the position response function, which takes the form
	\begin{equation}
		\tilde{\chi}(\omega) = \left( -\omega^2 + i\omega\tilde{\Gamma}_v^+(\omega) \right)^{-1}
		\label{eq_def_Response_fun}
	\end{equation}
	with $\tilde\Gamma_v^+(\omega)$ being the half sided Fourier transform of the friction kernel defined via
	\begin{equation}
		\tilde{\Gamma}_v^+(\omega) = \int_{0}^{\infty}e^{-i\omega t} \Gamma_v(t) dt \,.
		\label{eq_def_halfside_kernel_transform}
	\end{equation}
	The Fourier transform of the position correlation function $C_{xx}(t) = \langle x(0) x(t)\rangle $ can be written as
	\begin{equation}
		\tilde{C}_{xx}(\omega) = B\tilde{\chi}(\omega) \tilde{\Gamma}_R(\omega) \tilde{\chi}(-\omega) \,,
		\label{eq_ft_pos_corr}
	\end{equation}
	where we made use of 
	\begin{equation}
		\begin{split}
			\label{eq_ft_vacf}
			&\tilde{C}_{xx}(\omega) = \int_{-\infty}^{\infty} dt e^{-i\omega t}\langle x(0)x(t)  \rangle \\
			&= \int_{-\infty}^{\infty}e^{-i\omega t} dt \int_{\infty}^{\infty}e^{i\omega t} \frac{d\omega}{2\pi} \int_{-\infty}^{\infty} \frac{d\omega'}{2\pi} 
			\langle \tilde{x}(\omega) \tilde{x}(\omega')  \rangle
		\end{split}
	\end{equation}
	and
	\begin{equation}
		\label{eq_ft_force_corr}
		\langle  \tilde{F}_R(\omega) \tilde{F}_R(\omega') \rangle = 2\pi B \delta(\omega+\omega')\tilde{\Gamma}_R(\omega')\,.
	\end{equation}
	Equation \ref{eq_ft_pos_corr} can be rewritten as
	\begin{equation}
		\tilde{C}_{xx}(\omega) = \frac{B}{i\omega} \left( \tilde{\chi}(\omega) - \tilde{\chi}(-\omega) \right) \frac{\tilde{\Gamma}_R(\omega)}{\tilde{\Gamma}_v(\omega)} \,,
	\end{equation}
	which leads to the MSD
	\begin{equation}
		\begin{split}
			C_{\textrm{MSD}}(t) &= 2 (C_{xx}(0) - C_{xx}(t))\\
			&= B \int_{-\infty}^{\infty} \frac{d\omega}{\pi} \frac{e^{i\omega t} - 1}{i\omega}\left( \tilde{\chi}(\omega) - \tilde{\chi}(-\omega) \right) \frac{\tilde{\Gamma}_R(\omega)}{\tilde{\Gamma}_v(\omega)}
		\end{split}\,.
		\label{eq_msd_integral}
	\end{equation}
	For the decaying oscillation model friction kernel of eq. (7) and using the effective form $\Gamma(\vert t\vert)=\Gamma_R(t)=\Gamma_v(\vert t \vert)$ one obtains the half sided Fourier transform of the kernel as
	\begin{equation}
		\tilde{\Gamma}_+(\omega) =\int_{0}^{\infty}  \left(2a\delta(t)+be^{-t/\tau}\cos(\Omega t)\right)e^{-i\omega t} dt = a + b \left(\frac{i \omega + \frac{1}{\tau}}{\Omega^2 + (i\omega + \frac{1}{\tau})^2} \right)\,.
		\label{eq_fourier_half_kernel}
	\end{equation}
	Inserting the result of eq. \eqref{eq_fourier_half_kernel} into eq. \eqref{eq_def_Response_fun} we find the response function
	\begin{equation}
		\tilde{\chi}(\omega) = \left( -\omega^2 + i\omega a + i\omega b \left(\frac{i \omega + \frac{1}{\tau}}{\Omega^2 + (i\omega + \frac{1}{\tau})^2} \right) \right)^{-1} \,,
		\label{eq_Response_model}
	\end{equation} 
	which by inserting into eq. \eqref{eq_msd_integral} leads to
	\begin{equation}
		C_{\textrm{MSD}}(t) = B \int_{-\infty}^{\infty} \frac{d\omega}{\pi} \frac{(e^{i\omega t} - 1) (k_1 + k_2\omega^2 + k_3\omega^4)}{\omega^2(c_1 + c_2 \omega^2 + c_3 \omega^4 +\tau^4\omega^6)} \,,
		\label{eq_msd_int_inserted}
	\end{equation}
	where the constants $c_i$ and $k_i$ are given by
	\begin{eqnarray}
		c_1 &=& (a + b \tau + a \Omega^2 \tau^2)^2 \label{eq_c1_msd_int}\\
		c_2 &=& (1 + 2 (a^2 - b + \Omega^2) \tau^2 + 
		2 a b \tau^3 + (-2 a^2 \Omega^2 + (b + \Omega^2)^2) \tau^4) \\
		c_3 &=& \tau^2 (2 + (a^2 - 2 (b + \Omega^2)) \tau^2) \\
		k_1 &=& -2 (1 + \Omega^2 \tau^2) (a + b \tau + a \Omega^2 \tau^2) \\
		k_2 &=& -2 \tau^2 (b \tau + a (2 - 2 \Omega^2 \tau^2)) \\
		k_3 &=& -2 a \tau^4  \,.
		\label{eq_constants_msd_int}
	\end{eqnarray}
	Interpreting the integrand in eq. \eqref{eq_msd_int_inserted} as a sum of the three terms each proportional to $k_i$, we see that all terms have the same poles, where the term proportional to $k_1$ has an additional double pole at $\omega=0$. The solutions of $\omega^2$ to the equation
	\begin{equation}
		c_1 + c_2 \omega^2 + c_3 \omega^4 +\tau^4\omega^6 = \tau^4 (\omega^2-\omega_1^2)(\omega^2-\omega_2^2)(\omega^2-\omega_3^2)=0\,,
		\label{eq_poles_omegai_def}
	\end{equation}
	define the remaining poles, which we denote by $\pm\sqrt{\omega_i^2}$ and which we compute by  numerically solving eq. \eqref{eq_poles_omegai_def}.
	Next we use the partial fraction decompositions
	\begin{eqnarray}
		\hspace{-4mm}
		\frac{1}{\omega^2\prod_{i=1}^{3}(\omega^2-\omega_i^2)} &=& -\frac{1}{\omega^2\prod_{i=1}^{3}\omega_i^2} + \sum_{i=1}^{3} \frac{1}{\omega_i^2 (\omega^2-\omega_i^2) \prod_{j\neq i}(\omega_i^2-\omega_j^2)}
		\\
		\frac{1}{\prod_{i=1}^{3}(\omega^2-\omega_i^2)} &=&  \sum_{i=1}^{3} \frac{1}{(\omega^2-\omega_i^2)\prod_{j\neq i}(\omega_i^2-\omega_j^2)}
		\\
		\frac{\omega^2}{\prod_{i=1}^{3}(\omega^2-\omega_i^2)} &=&  \sum_{i=1}^{3} \frac{\omega_i^2}{(\omega^2-\omega_i^2)\prod_{j\neq i}(\omega_i^2-\omega_j^2)}
		\label{eq_partial_fraction_decomp}
	\end{eqnarray}
	to rewrite the fraction of eq. \eqref{eq_msd_int_inserted} as a sum of terms proportional to $(\omega^2-\omega_i^2)^{-1}$ and one term proportional to $\omega^{-2}$. Using the solution of the integrals
	\begin{eqnarray}
		\int_{-\infty}^{\infty}\frac{e^{i\omega t} -1}{\omega^2-\omega_i^2}d\omega  &=& \frac{\pi (e^{-\sqrt{-\omega_i^2}t} - 1)}{\sqrt{-\omega_i^2}}\\
		\int_{-\infty}^{\infty}\frac{ e^{i\omega t} - 1}{\omega^2}d\omega &=& -\pi t
		\label{eq_integrals_Roots}
	\end{eqnarray}
	for $t>0$ with the condition $\text{Re}(\omega_i^2)<0 \lor \text{Im}(\omega_i^2)\neq0 \land \text{Im}(\sqrt{\omega_i^2})\neq0$, we can rewrite the integral of the MSD eq. \eqref{eq_msd_int_inserted} as
	\begin{equation}
		C_{\textrm{MSD}}(t) = \frac{B}{\tau^4} \left( \frac{k_1 t}{\omega_1^2 \omega_2^2 \omega_3^2} + \sum_{i=1}^{3}\frac{e^{-\sqrt{-\omega_i^2} t} - 1}{\sqrt{-\omega_i^2} \prod_{j\neq i} (\omega_i^2-\omega_j^2) } \left[ \frac{k_1}{\omega_i^2} + k_2 + k_3 \omega_i^2 \right] \right) \,.
		\label{eq_msd_final_Res}
	\end{equation}
	The VACF can be computed from the MSD by using eq. (23) as
	\begin{equation}
		C_{vv}(t) = \frac{B}{2\tau^4} \left( \sum_{i=1}^{3}\frac{\sqrt{-\omega_i^2} e^{-\sqrt{-\omega_i^2} t}}{ \prod_{j\neq i} (\omega_i^2-\omega_j^2) } \left[ \frac{k_1}{\omega_i^2} + k_2 + k_3 \omega_i^2 \right] \right) \,.
		\label{eq_vacf_final_Res}
	\end{equation}
	
	The calculation of the MSD for the friction kernel of eq. (26) proceeds in the same way as shown for the kernel eq. (7). The response function then takes the slightly different form
	\begin{equation}
		\tilde{\chi}(\omega) = \left( -\omega^2 + i\omega a + i\omega b \left(\frac{i \omega + \frac{2}{\tau}}{\Omega^2 + (i\omega + \frac{1}{\tau})^2}\right) \right)^{-1} \,.
		\label{eq_Response_markov_embedding}
	\end{equation}
	The integral eq. \eqref{eq_msd_integral} resulting in the MSD can be written in the same form as before shown in eq. \eqref{eq_msd_int_inserted}, where the constants $c_1$, $c_2$, $k_1$, $k_2$ take different values of
	\begin{eqnarray}
		c_1 &=& (a + 2b \tau + a \Omega^2 \tau^2)^2 \label{eq_c1_msd_int_markov}\\
		c_2 &=& (1 + 2 (a^2 - 3b + \Omega^2) \tau^2 + 
		(-2 a^2 \Omega^2 + (b + \Omega^2)^2) \tau^4) \\
		k_1 &=& -2 (1 + \Omega^2 \tau^2) (a + 2b \tau + a \Omega^2 \tau^2) \\
		k_2 &=& 4 a \tau^2  (\Omega^2 \tau^2 - 1) \,.
		\label{eq_constants_msd_int_markov}
	\end{eqnarray}
	Thus, the MSD of the GLE eq. (1) with the friction kernel eq. (26) can be described by the same form eq. \eqref{eq_msd_final_Res} as for the friction kernel of eq. (7), where only the constants $c_i$ and $k_i$ have slightly different values as seen by comparing eqs. \eqref{eq_c1_msd_int}-\eqref{eq_constants_msd_int} to eqs. \eqref{eq_c1_msd_int_markov}-\eqref{eq_constants_msd_int_markov}. 
	
	\section{Localization noise fit}\label{sec_fit_localization_SI}
\begin{figure*}
		\centering
		\includegraphics[width=\textwidth]{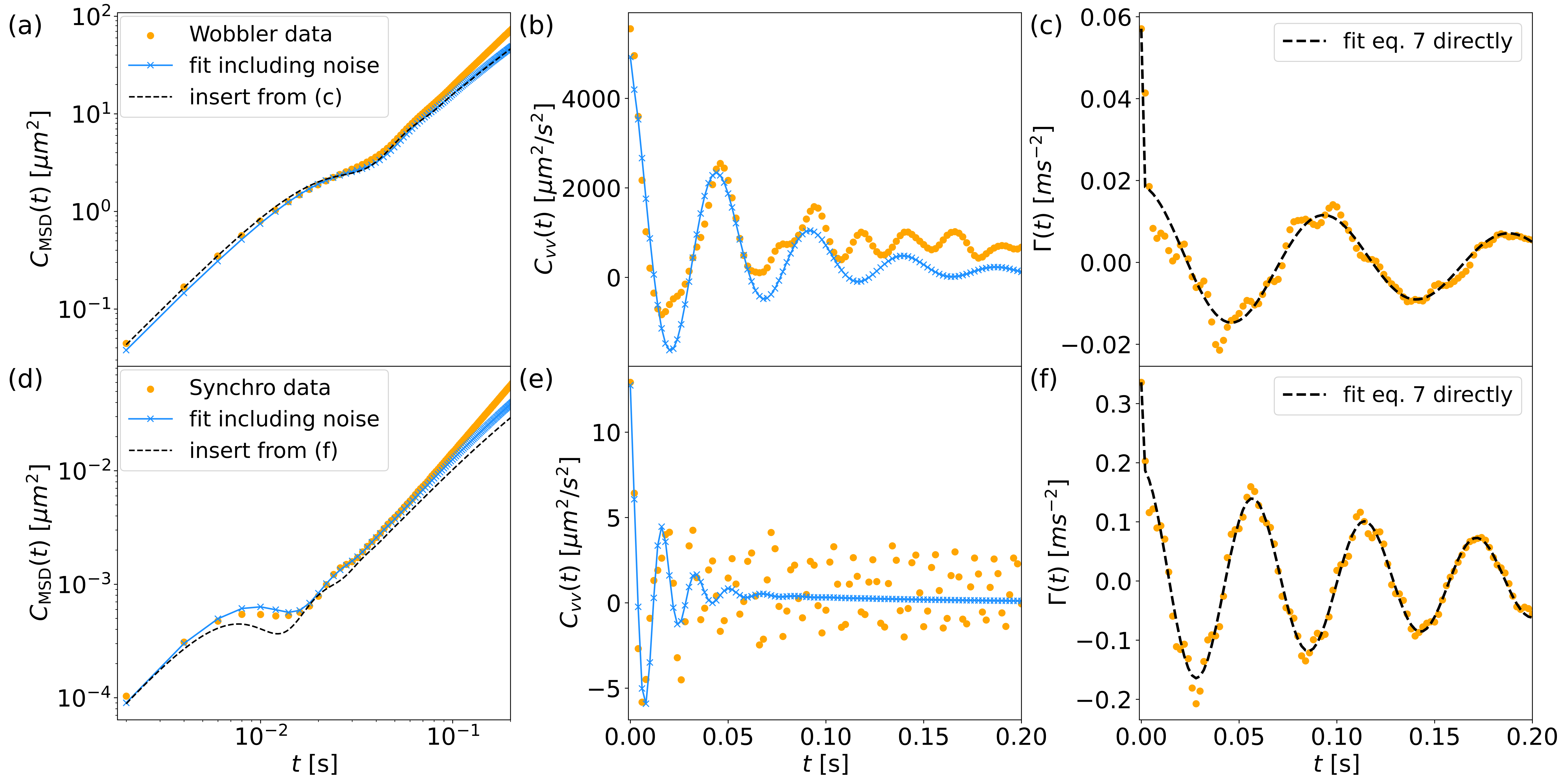}
		\caption{Two example cells for  which the  localization-noise fit to VACF does not agree well with the data. The MSD $C_{\textrm{MSD}}(t)$ (eq. (6)), VACF $C_{vv}(t)$ (eq. (3)) and friction kernel $\Gamma(t)$ extracted by eq. (15) of a single (a)-(c) wobbler and (d)-(f) synchro are displayed as orange dots. 
			Blue crosses in the VACF figures represent the results from the localization noise fit according to eq. (24)  and the blue crosses in the MSD figures result from inserting the fit parameters of the VACF into eq. (22).
			The black dashed line in (c) and (f) is the direct fit of eq. (7) to the extracted kernels, which is  used to obtain parameters for the cell classification, see Methods Sec. 4.6. The dashed black lines in the MSD figures result from inserting the fitting result of the friction kernel (dashed lines in (c), (f) respectively) and the mean squared velocity $B=C_{vv}(0)$ into the analytical solution of the MSD eq. \eqref{eq_msd_final_Res}.}
		\label{fig_bad_fits}
	\end{figure*}

	The fitting procedure including finite time step and localization noise described in Sec. 4.7 of the main text does sometimes not converge to results that agree well with the VACF of the data. In Fig. \ref{fig_bad_fits} we show two examples for which the fit including the localization noise (Methods eq. (24)) does not agree well with the VACF data.
	For 14 out of the 59 cells, the fits of the VACF using eq. (25) do not converge to a stable set of parameters. Still the direct fit of the friction kernel model eq. (7) agrees perfectly with the extracted friction kernels, which is why we use this direct fit to extract parameters, that are later used in the cluster analysis. 
	
	A typical size of the localization noise width resulting from the fits is ${\sigma_{\textrm{loc}}\approx 0.02\,\mu m}$.
	Considering the resolution of the microscope of roughly ${\sim 0.5\,\mu m}$ and approximating the area of a cell as roughly ${25\pi\,\mu m^2}$, one can estimate the number of pixels per cell as $100\pi$. Thus, the error of the mean position of a cell evaluated by all pixels is estimated by ${0.5\mu m/\sqrt{100\pi}\approx0.03\,\mu m}$, which is very close
	to the fitted  localization noise width of ${\sigma_{\textrm{loc}}\approx 0.02\,\mu m}$. For a typical velocity around $v=100\,\mu m/s$, the displacement during one time step $\Delta=0.002\,s$ is $0.2\,\mu m$. Therefore, the localization noise accounts for $\sim 10\%$ of the cell displacement. This relatively small localization noise explains why the direct fit of the model eq. (7) to the extracted friction kernels works so well and the inclusion of localization noise effects is not necessary for our data set.
	
	\section{Cluster analysis in lower dimensions}\label{sec_cluster_B_pca}
\begin{figure}
		\centering
		\includegraphics[width=\textwidth]{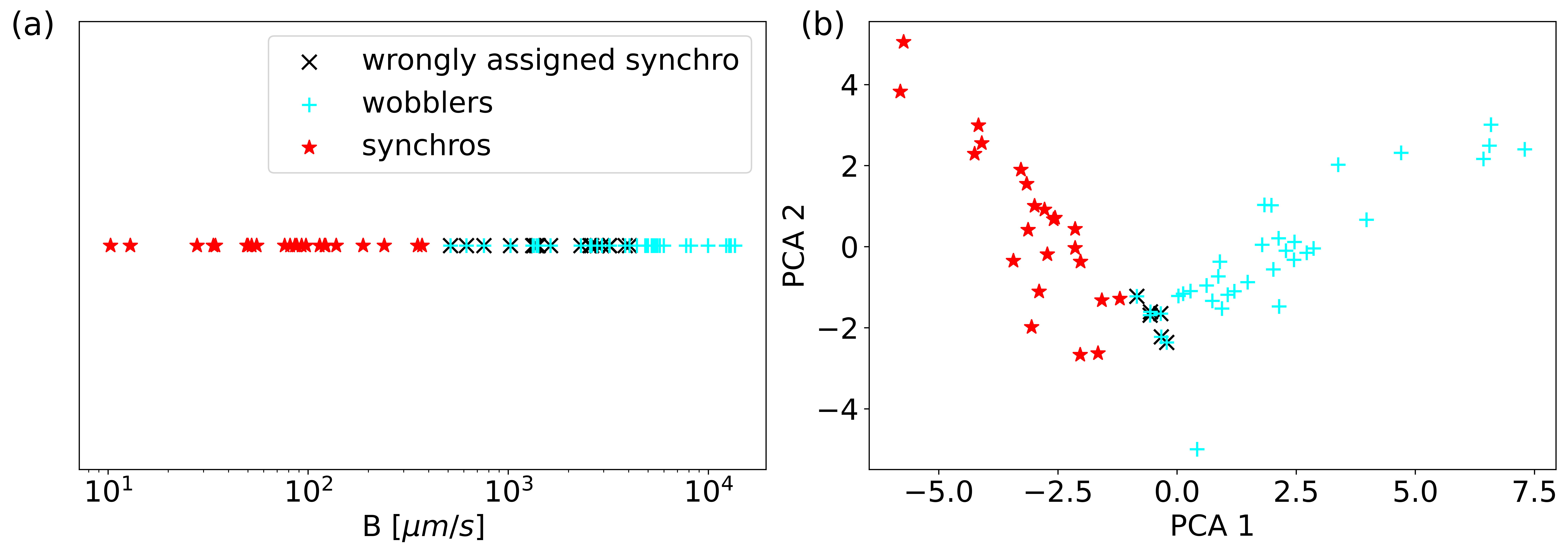}
		\caption{(a) Mean squared velocities $B$  for all cells;
			a cluster analysis solely based on B  leads to the wrong assignment of 18 wobblers as synchros, indicated by black crosses. (b) Projection of the five dimensional extracted parameters shown in Fig. 5 on the first two components of the PCA, which  explain $61\,\%$ and $22\,\%$ of the total parameter variance, respectively. The black crosses indicate  six wobbler cells that are wrongly assigned as synchros when applying the cluster analysis only to the shown two PCA components.}
		\label{fig_pca_msqv}
	\end{figure}

	Since the wobblers are found to exhibit higher average speeds than the synchros, an intuitive and simple approach would be to distinguish the cells solely by their mean squared velocity $B$. Applying the cluster analysis described in Sec. 4.8 of the Methods in one dimension for $B$, results in a assignment agreeing only to $69\,\%$ with the full classification result, 
	see Fig. \ref{fig_pca_msqv}a. Especially slow wobblers are wrongly assigned as synchrose,  as indicated in Fig. \ref{fig_pca_msqv}a by the black crosses.
	
	As discussed in the main text, high-dimensional data sets are often analyzed with principal component analysis (PCA), which determines the directions explaining most of the data variance. Applying a PCA to the five dimensional parameter space of the extracted friction kernel parameters eq. (7) and $B$ shown in Fig. 5, we find the first two components of the PCA to explain $83\,\%$ of the total parameter variance. Here, we use the PCA tool implemented in \textit{sklearn} for python. The normalized vectors of the shown two PCA components in the order ${(a,\hspace{1mm} b,\hspace{1mm} \tau,\hspace{1mm} \Omega,\hspace{1mm} B)}$ are given by ${(-0.49,\hspace{1mm} -0.29,\hspace{1mm}  0.16,\hspace{1mm} -0.01,\hspace{1mm}  0.81 )}$ and ${(0.63,\hspace{1mm} 0.41,\hspace{1mm}  -0.5,\hspace{1mm} 0.05,\hspace{1mm}  0.43 )}$, respectively. This indicates, that no parameter alone can describe the complete variance and therefore no parameter alone can explain all the differences between synchros and wobblers. The parameters most important for the variance and discriminability of the single CR cells are $a$, $b$ and $B$. However, the distinction into the two groups of wobblers and synchros by our cluster analysis (Sec. 4.8 in the Methods) works the best using the complete five dimensional set of parameters. Applying the cluster analysis to the first two PCA components results in an accuracy of $90\,\%$, as shown in Fig. \ref{fig_pca_msqv}b. Six wobblers are assigned to belong to the cluster of synchros (defined by the classification of \cite{mondal_strong_2021} that agrees with our cluster analysis of the complete parameter set Fig. 5) as they lie in the transition area between the two clusters and are indicated by the  black crosses in Fig. \ref{fig_pca_msqv}b.

	\section{Comparing friction kernel expressions eq. (7) and eq. (26)}\label{sec_similar_kernel}
\begin{figure}
		\centering
		\includegraphics[width=\textwidth]{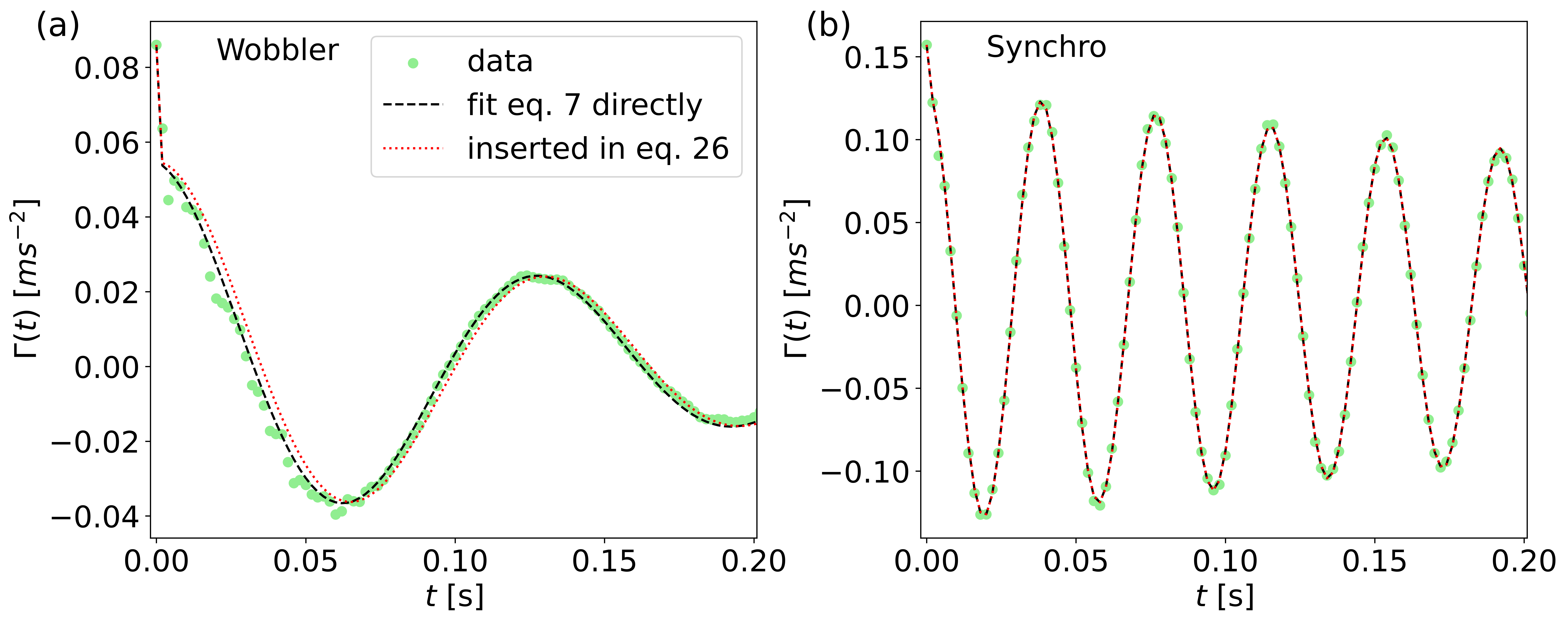}
		\caption{Friction kernel $\Gamma(t)$ extracted by eq. (15) of (a) a single wobbler and (b) single synchro shown as green data points. The black line shows the direct fit of eq. (7) to the data points and the red dotted line results from inserting the fitted parameters from the black dashed line into eq. (26) derived from the Hamiltonian eq. (27) in Sec. \ref{sec_markov_embedding}. The agreement of the black dashed line and the red dotted line means that in the parameter range exhibited by the CR cells, the two kernels eq. (7) and eq. (26) describe the data equally well.}
		\label{fig_compare_kernels}
	\end{figure}

	The friction kernel of eq. (7) in the main text used to extract parameters from the data is very similar to the friction kernel eq. (26), which is derived in Sec. \ref{sec_markov_embedding} from a system of harmonically coupled particles. The term $1/(\tau\Omega)$ in front of the sine  in eq. (26)
	becomes small when the decay time $\tau$ is longer than the oscillation period $1/\Omega$. Several oscillations occur before the friction kernel has decayed for the CR data in Fig. 3 in the main text and Fig. \ref{fig_compare_kernels}, thus, the term $1/(\tau\Omega)$ is indeed  small and the two friction kernel models have very similar shapes, 
	as we show in Fig. \ref{fig_compare_kernels}. Here, we insert the extracted parameters from the friction kernel eq. (7) of two example cells into eq. (26),
	the deviations are seen to be very small.
	
	\section{Exemplary non-equilibrium model describing CR motion}\label{sec_noneq_mapping_example}
	The effective friction kernel $\Gamma(t)$ given by eq. (26) leads to the Fourier transform
	\begin{align}
		\tilde{\Gamma}_+(\omega) &= \int_{0}^{\infty}  \left( 2a\delta(t) + be^{-t/\tau}\left(\cos(\Omega t) + \frac{1}{\tau\Omega}\sin(\Omega t)\right) \right)e^{-i\omega t} dt \nonumber \\
		&= a + b \left(\frac{i \omega + \frac{2}{\tau}}{\Omega^2 + (i\omega + \frac{1}{\tau})^2}\right) \label{eq_fourier_half_kernel_emb}\,.
	\end{align}
	Inserting eq. \eqref{eq_fourier_half_kernel_emb} into eq. (11) yields the Fourier-transformed VACF 
	\begin{equation}
		\begin{split}
			&\frac{\tilde{C}_{vv}(\omega)}{B} = \bigg( 2 a + 2 b \tau + 4 a \Omega^2 \tau^2 + 2 b \Omega^2 \tau^3 + 2 a \Omega^4 \tau^4 + 
			4 i b \tau^2 \omega \\
			&+ (4 a \tau^2 + 2 b \tau^3 - 
			4 a \Omega^2 \tau^4) \omega^2 + 2 a \tau^4 \omega^4 \bigg)/ \\
			&\bigg( a + 2 b \tau + a \Omega^2 \tau^2)^2 + (1 + 2 (a^2 - 3 b + \Omega^2) \tau^2 + (-2 a^2 \Omega^2 + (b + \Omega^2)^2) \tau^4) \omega^2 \\
			&+ \tau^2 (2 + (a^2 - 2 (b + \Omega^2)) \tau^2) \omega^4 + \tau^4 \omega^6 \bigg) \\
		\end{split}\,.
		\label{eq_vacf_ft_markov}
	\end{equation}
	In order to find a non-equilibrium model, that leads to the same VACF as the kernel in eq. (26), we assume a specific form of the velocity friction kernel
	\begin{equation}
		\Gamma_v(t) = 2a_v\delta(t)
		\label{eq_gammav_delta_example_SI}
	\end{equation}
	which leads to the Fourier transform ${\tilde{\Gamma}_v^+(\omega)=a_v}$. We insert this together with the result from eq. \eqref{eq_vacf_ft_markov} into eq. (12) and solve for $\tilde{\Gamma}_R(\omega)$.
	By Fourier back transformation the result  can be written in the form 
	\begin{equation}
		\Gamma_R(t) = \int_{-\infty}^{\infty} \frac{d\omega}{2\pi} e^{i\omega t} \frac{ k_0 +k_1\omega + k_2\omega^2 +k_3\omega^3 + k_4\omega^4 + k_6\omega^6}{c_0 + c_1 \omega^2 + c_2 \omega^4 +c_3\omega^6}
		\label{eq_gammar_mapping_example_SI}
	\end{equation}
	with the constants $c_i$ and $k_i$ being
	\begin{eqnarray}
		c_0 &=& (a + 2b \tau + a \Omega^2 \tau^2)^2 \nonumber\\
		c_1 &=& 1 + 2 (a^2 - 3 b + \Omega^2) \tau^2 + (-2 a^2 \Omega^2 + (b + \Omega^2)^2) \tau^4 \nonumber\\
		c_2 &=& \tau^2 (2 + (a^2 - 2 (b + \Omega^2)) \tau^2) \nonumber\\
		c_3 &=& \tau^4 \nonumber\\
		k_0 &=& 2 a a_v^2 + 2 a_v^2 b \tau + 4 a a_v^2 \Omega^2 \tau^2 + 2 a_v^2 b \Omega^2 \tau^3 + 
		2 a a_v^2 \Omega^4 \tau^4 \nonumber\\
		k_1 &=& 4 a_v^2 b \tau^2 \nonumber\\
		k_2 &=& 2 (b \tau (1 + a_v^2 \tau^2 + \Omega^2 \tau^2) + 
		a ((1 + \Omega^2 \tau^2)^2 + a_v^2 (2 \tau^2 - 2 \Omega^2 \tau^4))) \nonumber\\
		k_3 &=& 4 b \tau^2 \nonumber\\
		k_4 &=& 2 \tau^2 (b \tau + a (2 + a_v^2 \tau^2 - 2 \Omega^2 \tau^2)) \nonumber\\
		k_6 &=& 2a \tau^4 \nonumber\,.
	\end{eqnarray}
	We solve this Fourier transform by performing the same steps as for the calculation of the integral \eqref{eq_msd_int_inserted} derived in Sec. \ref{sec_msd_ana}, where $\omega_i^2$ are the roots of the denominator of eq. \eqref{eq_gammar_mapping_example_SI} in $\omega^2$, which we obtain numerically, and we additionally use
	\begin{equation}
		\int_{-\infty}^{\infty} \frac{\omega e^{i\omega t} }{\omega^2-\omega_i^2}d\omega  = i \pi e^{-\sqrt{-\omega_i^2}t} \,.
		\label{eq_decomposition_odd_integral}
	\end{equation}
	The friction kernel in time domain is then retrieved as
	\begin{equation}
		\begin{split}
			\Gamma_R(t) = 2a\delta(t) + \frac{1}{2 c_3}\sum_{i=1}^{3} \bigg(\frac{e^{-\sqrt{-\omega_i^2} t}}{\sqrt{-\omega_i^2} \prod_{j\neq i} (\omega_i^2-\omega_j^2) }
			\left[ k_0 + k_2 \omega_i^2 + k_4 (\omega_i^2)^2 +k_6 (\omega_i^2)^3 \right]  \\
			-\frac{ e^{-\sqrt{-\omega_i^2} t}}{\prod_{j\neq i} (\omega_i^2-\omega_j^2) } \left[ k_1 + k_3 \omega_i^2\right] \bigg).
		\end{split}\,
		\label{eq_gammar_mapping_result}
	\end{equation}
	This exemplary mapping shows, that a simple Markovian friction kernel given by eq. \eqref{eq_gammav_delta_example_SI}
	combined with a  random force correlation function that contains additional oscillating components,given by eq. \eqref{eq_gammar_mapping_result}, lead to the same correlation function as described by  the GLE 
	with the effective friction kernel given by eq. (26). We have thus derived one possible non-equilibrium model that describes CR cell motion.
	
	Furthermore, this non-equilibrium model with $\Gamma_R(t)$ given by eq. \eqref{eq_gammar_mapping_result} and $\Gamma_v(t)$ given by eq. \eqref{eq_gammav_delta_example_SI} corresponds to  the coupled system of differential equations
	\begin{align}
		\dot{x}(t) &= v(t)\label{eq_emb_noneq_0}\\
		\dot{v}(t) &= -a_v v(t) + \xi_x(t) + \sum_{i=1}^{3} F_i(t) \label{eq_emb_noneq_1}\\
		\dot{F}_i(t) &= -\frac{1}{\tau_i} \left( F_i(t) - \xi^R_i(t) \right)\label{eq_emb_noneq_2}\,,
	\end{align}
	with $\langle \xi_x(0)\xi_x(t) \rangle = 2aB\delta(t)$ and $\langle \xi^R_i(0)\xi^R_j(t) \rangle = 2 a^R_i B \delta_{ij} \delta(t)$.\\
	Solving eq. \eqref{eq_emb_noneq_2} for $F_i(t)$ leads to
	\begin{equation}
		F_i(t) = -\frac{1}{\tau_i}\int_{0}^{t} e^{-(t-t')/\tau_i} \xi^R_i(t') dt'\,,
	\end{equation}
	which yields $\langle F_i(0)F_i(t) \rangle = B\frac{a^R_i}{\tau_i}e^{-t/\tau_i}$. Now defining the random force as $F_R(t) = \xi_x(t) + \sum_{i=1}^{3} F_i(t)$, we can write eq. \eqref{eq_emb_noneq_1} as the GLE eq. (1) with
	\begin{equation}
		\Gamma_R(t) = 2a\delta(t) + \sum_{i=1}^{3} \frac{a^R_i}{\tau_i}e^{-t/\tau_i}
	\end{equation}
	and $\Gamma_v(t) = 2a_v\delta(t)$.
	Thus, the system of eqs. \eqref{eq_emb_noneq_0}-\eqref{eq_emb_noneq_2} is equivalent to the GLE eq. (1) with $\Gamma_R(t)$ given by eq. \eqref{eq_gammar_mapping_result} and $\Gamma_v(t)$ given by eq. \eqref{eq_gammav_delta_example_SI}, when the parameters are given by
	\begin{align}
		\tau_i &= \frac{1}{\sqrt{-\omega_i^2}}\\
		a^R_i &= \frac{1}{-2c_3 \omega_i^2 \prod_{j\neq i} (\omega_i^2-\omega_j^2)} \left[ k_0 + k_2 \omega_i^2 + k_4 (\omega_i^2)^2 +k_6 (\omega_i^2)^3 - \sqrt{-\omega_i^2} (k_1 + k_3 \omega_i^2)  \right]\,.
	\end{align}

	\section{Information on the cell orientation is  contained in the  cell-center trajectories}\label{sec_orientation_SI}
\begin{figure*}
		\centering
		\includegraphics[width=\textwidth]{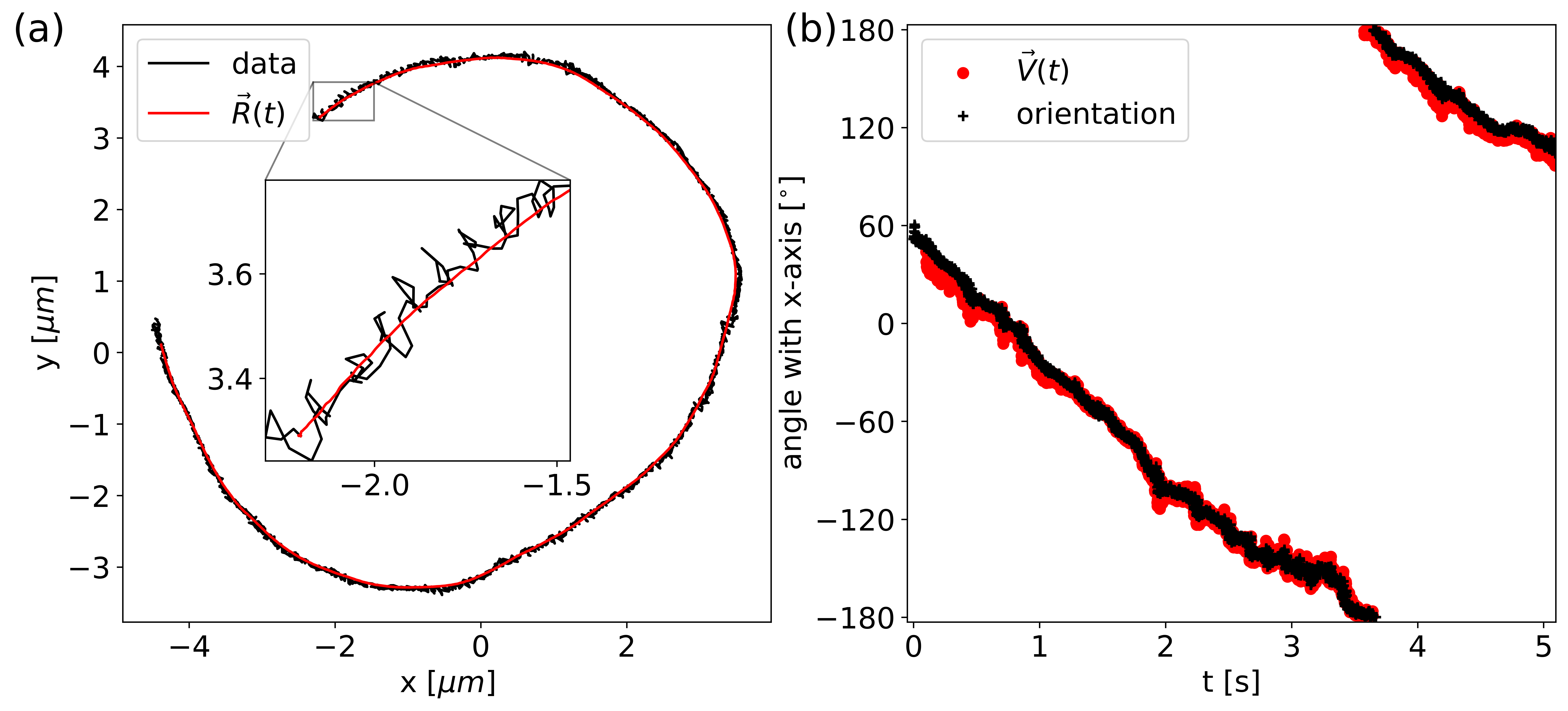}
		\caption{(a) Trajectory  of a synchro shown in black,  the running average position defined by eqs. \eqref{eq_orientation_v}, \eqref{eq_orientation_r} with a decay time $T=0.1\,s$ is shown in red. The inset in the middle shows a zoom into the first $0.2\,s$ of the trajectory. (b) The angle between the x-axis and the cell orientation directly extracted from video data is shown for the trajectory in (a) as black crosses. The angle between the x-axis and the running-average velocity $\vec{V}(t)$ determined by eq. \eqref{eq_orientation_v}  is shown in red.}
		\label{fig_orientation}
	\end{figure*}

	The mean direction of motion,  exponentially weighted  over the past trajectory,  is defined by
	\begin{equation}
		\vec{V}(t)=\frac{1}{T}\int_{-\infty}^{t}e^{-(t-t')/T }\vec{v}(t')dt' \hspace{3mm},
		\label{eq_orientation_v}
	\end{equation}
	where $T$ is the exponential decay time  and $\vec{v}(t)=(v_x(t), v_y(t))$ is the two-dimensional velocity \cite{li2011dicty}.
	This running-average velocity vector $\vec{V}(t)$ points in the direction in which a cell has  moved in the past,
	the running-average position follows  as
	\begin{equation}
		\vec{R}(t) = \int_{0}^{t}\vec{V}(t')dt'\,.
		\label{eq_orientation_r}
	\end{equation}
	In Fig. \ref{fig_orientation}a we show the trajectory of a synchro and its running average position $\vec{R}(t)$ for $T=0.1\,s$, 
	$\vec{R}(t)$  resolves the mean direction of motion but not the back-and-forth motion on time scales below $T$. 
	From the circular path of the trajectory it follows that the mean direction of motion changes, 
	this is reflected in Fig. \ref{fig_orientation}b by the change of the angle of  $\vec{R}(t) $ with  the x-axis (red data).

	The orientation of cells is determined from the videos by fitting an ellipsoid to the cell perimeter and given by the main axis of the ellipsoid. Since CR cells move with their flagella in front, the orientation vector is defined to point towards the side where the flagella are anchored.

	The cell orientation vector coincides very well with the direction of the running average velocity $\vec{V}(t)$, as shown in Fig. \ref{fig_orientation}b. 
	Therefore, the information on the orientation of the CR cells is included in the cell trajectory $(v_x(t), v_y(t))$ via eq. \eqref{eq_orientation_v}. Our description of the CR motion by the GLE eq. (1) captures the correct behavior of the cell position and velocity over time, which is demonstrated by Figs. 4a-f. Since the cell orientational information
	is included in the trajectory, we conclude that the GLE model also correctly describes  the CR orientational dynamics.
	
	\section{Effects of smoothing trajectory data}\label{sec_smoothing_SI}
	Since the localization noise modifies the VACF and thereby the friction kernel that is extracted from the data, it is important to 
	smooth the data at a level that reduces the  noise without loosing detail. 
	For every data set, there is an optimal level of smoothing. 
	
\begin{figure*}
		\centering
		\includegraphics[width=\textwidth]{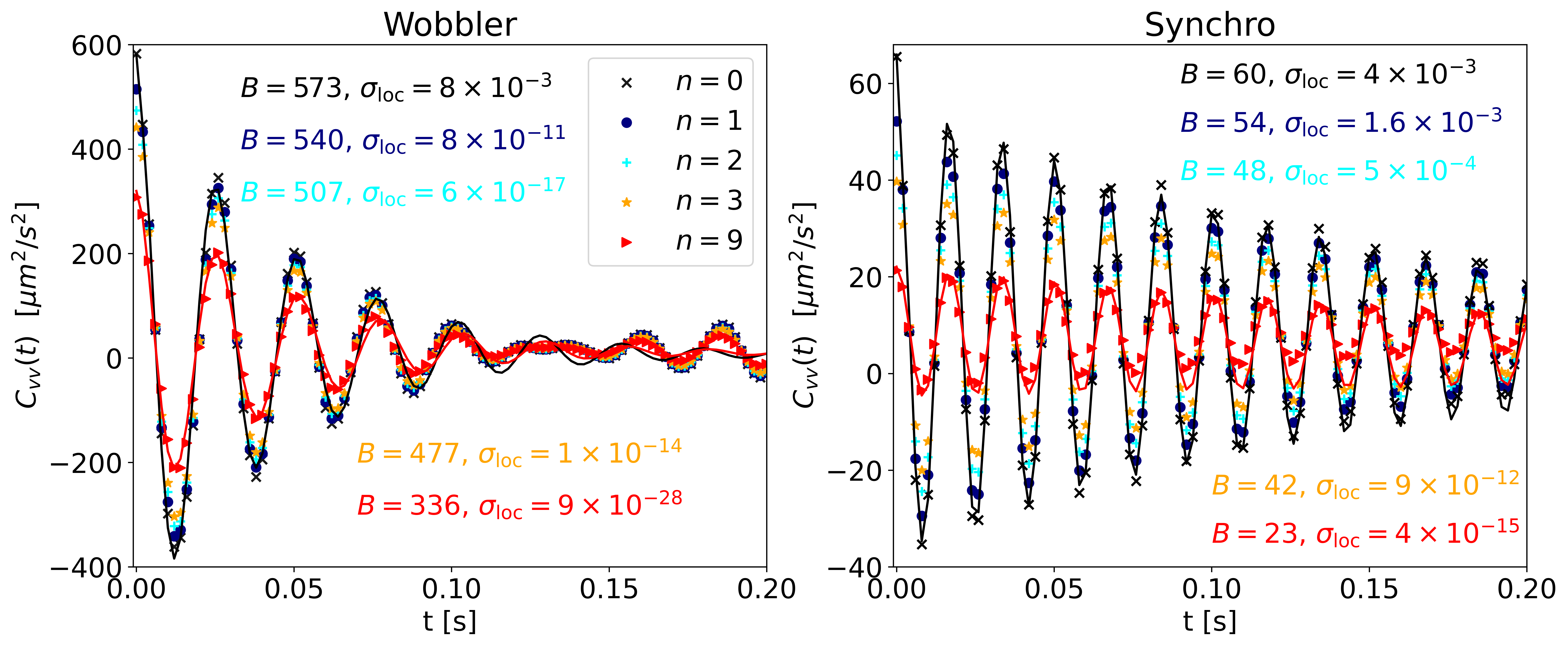}
		\caption{The VACF  $C_{vv}(t)$ of a wobbler (left) and  a synchro  (right) is shown for different smoothing iterations $n$ described by eq. \eqref{eq_smoothing_SI}, where $n=0$ denotes the original non-smoothed data. The mean-squared velocities $B$ in units of $\mu m^2/s^2$ and the fitted  localization noise width $\sigma_{\textrm{loc}}$ in units of $\mu m$ are both given in the color of the according smoothing iteration. The fitting result of the VACF including localization noise, described in Sec. \ref{sec_fit_localization_SI}, is shown as  solid lines for $n=0$ and $n=9$ in the respective color.}
		\label{fig_vacf_smoothing}
	\end{figure*}

\begin{figure*}
		\centering
		\includegraphics[width=\textwidth]{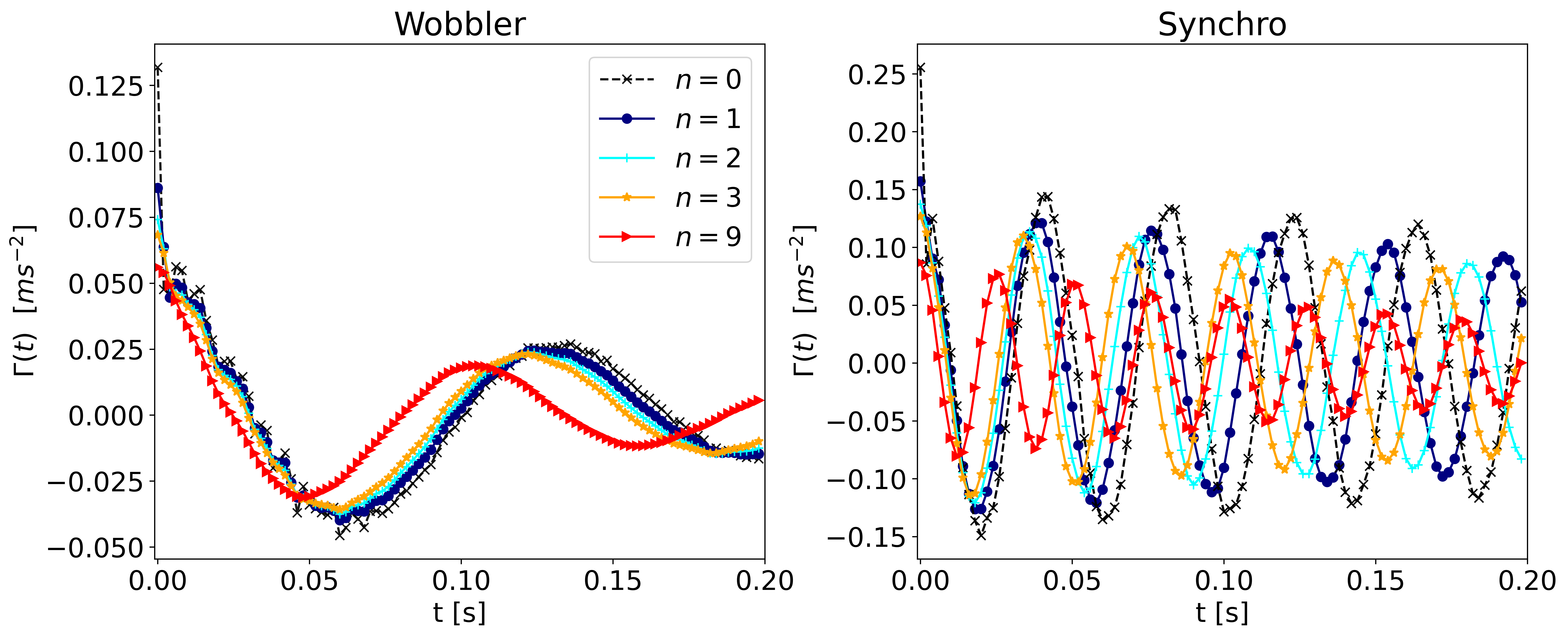}
		\caption{The  friction kernel $\Gamma(t)$ extracted according to  eq. (15)  
			from data at different smoothing iterations described by eq. \eqref{eq_smoothing_SI} is given for a wobbler on the left and for a synchro on the right. The solid lines connecting the data points are guides to  the eye.}
		\label{fig_kernel_smoothing}
	\end{figure*}

	A natural way of smoothing data is by sequentially averaging over neighboring data points, 
	which conserves the discretization  time step.
	Here, we use an iterative averaging of the position data over two consecutive points according to 
	\begin{equation}
		x_i^{n+1} = \frac{x_i^{n} + x_{i+1}^{n}}{2}\,,
		\label{eq_smoothing_SI}
	\end{equation}
	with $n$ being the number of smoothing iterations.
	We show the effect  of the smoothing level on the VACF and the  localization noise fit in Fig. \ref{fig_vacf_smoothing} 
	and the effect on the friction kernel in Fig. \ref{fig_kernel_smoothing}. For high smoothing iterations, the cell speed is underestimated,  which is reflected by a decreasing mean squared velocity $B$ for increasing $n$ in Fig. \ref{fig_vacf_smoothing}. 
	At the same time, the fitted localization noise strength $\sigma_{\textrm{loc}}$ decreases with increasing $n$. 
	This is reflected by the  smaller difference between the first two data points of the VACF. 
	
	
	The estimate of the localization noise strength for high smoothing iterations $n$ is not reliable anymore 
	because it becomes very small,	
	this explains why it is not  decreasing monotonically with growing $n$, see Fig. \ref{fig_vacf_smoothing}. 
	Since the first smoothing iteration $n=1$ simply adds two consecutive points according to eq. \eqref{eq_smoothing_SI}, 
	one in principle expects the localization noise width $\sigma_{\textrm{loc}}$ to be halved compared to the non-smoothed data. This indeed one can see  for the synchro shown in Fig. \ref{fig_vacf_smoothing}.
	
	Since the smoothing of the data decreases the amplitude of the VACF, as seen in Fig. \ref{fig_vacf_smoothing}, it consequently decreases the amplitudes of the friction kernel parameters  $a$ and $b$ and therefore at the same time it reduces the kernel frequency $\Omega$, as $\Omega$ exhibits a complex relation with the frequency of the VACF, which is explained in detail in Sec. \ref{sec_frequency_vacf_kernel}. Moreover, we find the dip in the friction kernel at the first time step to decrease for higher smoothing iterations shown in Fig. \eqref{fig_kernel_smoothing}, which indicates that the dip in the friction kernel originates from the localization noise. This is explained by  eq. (15), since the first point of the VACF is overestimated and the second point of the VACF is underestimated due to the localization noise, which then propagates into the friction kernel and leads to a dip at short times.
	
	The VACF of the smoothed data is less influenced by the localization noise, nevertheless, with every smoothing iteration, the deviation of the smoothed VACF  from the non-smoothed VACF  becomes larger, as shown in Fig. \ref{fig_vacf_smoothing}. Thus, we choose $n=1$ smoothing iterations for our data as a compromise between minimizing the effect of noise and keeping a good resolution that accurately describes the actual cell velocities.
	
	\section{Markovian embedding of the  friction kernel eq. (26)}\label{sec_markov_embedding}
	Here we derive  the equivalency of the system of coupled equations of motion
	\begin{eqnarray}
		\dot{x}(t) &=& v(t) \label{eq_markov_embedding_SI1}\\
		m \dot{v}(t) &=& -\gamma_x v + b m (y(t) - x(t)) + F_{Rx}(t) \label{eq_markov_embedding_SI2} \\
		\dot{y}(t) &=& v_y(t) \label{eq_markov_embedding_SI3}\\
		m_y \dot{v}_y(t) &=& -\gamma_y v_y + b m (x(t) - y(t)) + F_{Ry}(t)\label{eq_markov_embedding_SI4} \,,
		\label{eq_markov_embedding_SI}
	\end{eqnarray}
	describing  two harmonically coupled masses in a heat bath  based on the Hamiltonian eq. (27), to the GLE eq. (1) with the effective friction kernel of eq. (26) \cite{brunig2022time}.
	Here, the random forces have zero mean $\langle F_i(t) \rangle = 0$ and are delta correlated $\langle F_i(0)F_j(t)\rangle = 2\gamma_i k_BT \delta_{ij}\delta(t)$.
	We start by finding the solution for $\vec{z}(t)=(y,\dot{y})^T$ to the subsystem eqs. \eqref{eq_markov_embedding_SI3} and \eqref{eq_markov_embedding_SI4}, which is the solution to
	\begin{equation}
		\dot{\vec{z}}(t) = A\vec{z}(t) + \begin{pmatrix}
			0\\
			F_{Ry}(t)/m_y +\frac{bm}{m_y}x(t)
		\end{pmatrix}
		\label{eq_differential_vec}
	\end{equation}
	with
	\begin{equation}
		A = \begin{pmatrix}0 & 1\\
			-\frac{bm}{m_y} & -\frac{\gamma_y}{m_y}
		\end{pmatrix}
	\end{equation}
	and consequently
	\begin{equation}
		A^{-1} = \begin{pmatrix}-\frac{\gamma_y}{bm} & -\frac{m_y}{bm}\\
			1 & 0
		\end{pmatrix}\,.
	\end{equation}
	One finds the solution to eq. \eqref{eq_differential_vec} as
	\begin{equation}
		\vec{z}(t) = e^{A [t-t_0]}\vec{z}(t_0) + \int_{t_0}^{t}dt' e^{A[t-t']} \begin{pmatrix}
			0\\
			F_{Ry}(t')/m_y +\frac{bm}{m_y}x(t')
		\end{pmatrix}\,.
		\label{eq_solution_y}
	\end{equation}
	Performing a partial integration on the integral in eq. \eqref{eq_solution_y} for the term including $x(t')$ leads to
	\begin{equation}
		\begin{split}
			\vec{z}(t) = e^{A [t-t_0]}\vec{z}(t_0) + \int_{t_0}^{t}dt' A^{-1} e^{A[t-t']} \begin{pmatrix}
				0\\
				\frac{bm}{m_y}v(t')
			\end{pmatrix}\\
			- \left[ A^{-1}e^{A[t-t']} \begin{pmatrix}
				0\\
				\frac{bm}{m_y}x(t')
			\end{pmatrix}  \right]_{t_0}^t\\
			+ \int_{t_0}^{t}dt' e^{A[t-t']} \begin{pmatrix}
				0\\
				F_{Ry}(t')/m_y
			\end{pmatrix},
		\end{split} \,,
		\label{eq_y_after_partial_int}
	\end{equation}
	where the matrix exponential $e^{At}$ can be expressed in terms of the Eigenvalues of $A$
	\begin{equation}
		\lambda_{1,2} = -\frac{\gamma_y}{2m_y} \pm \omega_0
		\label{eq_eigenvalues_A}
	\end{equation}
	with 
	\begin{equation}
		\omega_0=\sqrt{\left(\frac{\gamma_y}{2m_y}\right)^2-\frac{bm}{m_y}}
		\label{eq_omega_zero}
	\end{equation}
	as
	\begin{equation}
		\begin{aligned}
			e^{At} &= \frac{1}{\lambda_2-\lambda_1}\begin{pmatrix}
				\lambda_2 e^{\lambda_1 t} - \lambda_1 e^{\lambda_2 t} &  e^{\lambda_2 t} - e^{\lambda_1 t}\\
				\lambda_1\lambda_2(e^{\lambda_1 t} - e^{\lambda_2 t}) & \lambda_2 e^{\lambda_2 t} - \lambda_1 e^{\lambda_1 t}
			\end{pmatrix}\\
			&= e^{-t \frac{\gamma_y}{2 m_y}}\begin{pmatrix}
				\cosh(\omega_0 t)+ \frac{\sinh(\omega_0 t)\gamma_y}{2 m_y\omega_0} &  \sinh(\omega_0 t)/\omega_0\\
				-\frac{\sinh(\omega_0 t)bm}{\omega_0 m_y} & \cosh(\omega_0 t) - \frac{\sinh(\omega_0 t)\gamma_y}{2 m_y\omega_0}
			\end{pmatrix}
		\end{aligned}
		\label{eq_exponential_matrix_A}
	\end{equation}
	consequently leading to
	\begin{equation}
		A^{-1}e^{At} = e^{-t \frac{\gamma_y}{2 m_y}}\begin{pmatrix}
			-\frac{\cosh(\omega_0 t)\gamma_y}{bm} + \frac{\sinh(\omega_0 t)}{\omega_0} (1-\frac{\gamma_y^2}{2b m m_y}) &  -\frac{\sinh(\omega_0 t)\gamma_y}{2b m\omega_0} - \frac{\cosh(\omega_0 t)m_y}{b m}\\
			\cosh(\omega_0 t) + \frac{\sinh(\omega_0 t)\gamma_y}{2m_y \omega_0}& \frac{\sinh(\omega_0 t)}{\omega_0}
		\end{pmatrix}\,.
		\label{eq_inva_times_expa}
	\end{equation}
	The solution for $y(t)$ can now be found as the first element of the vector $\vec{z}(t)$ from eqs. \eqref{eq_y_after_partial_int}-\eqref{eq_inva_times_expa} when assuming a stationary system, i.e. $t_0\rightarrow-\infty$ as
	\begin{equation}
		\begin{split}
			y(t) &= x(t) - \int_{-\infty}^{t}dt' v(t')e^{-\frac{(t-t')\gamma_y}{2m_y}} \left( \cosh(\omega_0(t-t')) + \frac{\gamma_y}{2m_y \omega_0}\sinh(\omega_0(t-t')) \right)  \\
			&+ \int_{-\infty}^{t}dt' F_{Ry}(t')e^{-\frac{(t-t')\gamma_y}{2m_y}} \frac{\sinh(\omega_0 (t-t'))}{m_y\omega_0}.
		\end{split}
		\label{eq_sol_yt_position}
	\end{equation}
	Inserting eq. \eqref{eq_sol_yt_position} into eq. \eqref{eq_markov_embedding_SI2}, one recovers the GLE eq. (1) for $x(t)$
	\begin{equation}
		m\ddot{x}(t) = -\int_{-\infty}^{t}dt' \bar{\Gamma}(t-t') +\bar{F}_R(t),
	\end{equation}
	where the bars indicate multiplication by the mass $\Gamma(t)=\bar{\Gamma}(t)/m$, $F_R=\bar{F}_R/m$ and
	the friction kernel takes the oscillatory form
	\begin{equation}
		\Gamma(t) =\frac{\gamma_x}{m}\delta(t) + b e^{-t \frac{\gamma_y}{2 m_y}} \left( \cosh(\omega_0 t) + \frac{\gamma_y}{2 m_y \omega_0} \sinh(\omega_0 t)\right)
		\label{eq_kernel_embedded}
	\end{equation}
	and the random force is explicitly given by
	\begin{equation}
		F_R(t) = \bar{F}_{Rx}/m + b \int_{-\infty}^{t}dt' \bar{F}_{Ry}(t')e^{-\frac{(t-t')\gamma_y}{2m_y}} \frac{\sinh(\omega_0 (t-t'))}{m_y\omega_0} \,.
		\label{eq_random_force_embedded}
	\end{equation}
	Computing the random force correlation, one finds
	\begin{equation}
		\langle F_R(0)F_R(t) \rangle = B \Gamma(t) \,.
	\end{equation}
	The friction kernel of eq. (26) in the Methods is equivalent to the friction kernel of eq. \eqref{eq_kernel_embedded} with the parameters given by
	\begin{eqnarray}
		a &=& \frac{\gamma_x}{m} \nonumber\\
		\tau &=& \frac{2m_y}{\gamma_y} \nonumber\\
		\Omega &=& i\omega_0 \nonumber \,.
	\end{eqnarray}

	\bibliography{cell_classification.bib}